\documentclass[twocolumn,trackchanges]{aastex7}

\newcommand{\feh}{\ensuremath{\left[{\rm Fe}/{\rm H}\right]}}
\newcommand{\msun}{\ensuremath{\,M_\Sun}}
\usepackage[T1]{fontenc} 
\usepackage[utf8]{inputenc} 
\usepackage{amssymb}
\usepackage{float}
\usepackage{placeins}
\usepackage{hyperref}
\usepackage{natbib}
\usepackage{booktabs}
\usepackage{amsmath}

\setcitestyle{notesep={ }}

\begin{document}


\title{High-Eccentricity Tidal Migration Driven by Secular Chaos in Wide-Binary Systems}

\correspondingauthor{Yurou Liu}

\author[0009-0007-9211-2884]{Yurou Liu} 
\affiliation{Department of Astronomy, Yale University, New Haven, CT 06511, USA}
\email{yurou.liu@yale.edu}

\author[0000-0002-5181-0463]{Hareesh Gautham Bhaskar}
\affiliation{Department of Astronomy, Indiana University, Bloomington, IN 47405, USA}
\email{hbhaskar@iu.edu}

\author[0000-0002-0376-6365]{Xian-Yu Wang}
\affiliation{Department of Astronomy, Indiana University, Bloomington, IN 47405, USA}
\email{xwa5@iu.edu}

\author[0000-0003-0412-9314]{Cristobal Petrovich}
\affiliation{Department of Astronomy, Indiana University, Bloomington, IN 47405, USA}
\email{cpetrovi@iu.edu}
 
\begin{abstract}
High-eccentricity tidal migration driven by a distant stellar companion offers a natural pathway for producing some hot Jupiters; yet, most theoretical work has relied on an idealized three-body configuration whose simplicity makes the problem especially tractable. In reality, many cold-Jupiter systems may host additional planets or substellar objects, whose interactions can dramatically alter the pathways to secularly excite extreme eccentricities.
We investigate how secular chaos can drive high-eccentricity tidal migration in hierarchical “3+1” systems—stellar binaries hosting a planet and an additional intermediate companion orbiting the primary star.
We show that the onset of secular chaos is regulated by the ratio of the von-Zeipel-Lidov-Kozai (ZLK) timescales of the inner and outer orbits $\mathcal{R}$.
When $\mathcal{R}\sim 0.5-2$, most systems can undergo migration even when their mutual inclinations remain modest—below the $39.2^\circ$ critical angle for ZLK oscillations---with diffusion timescales spanning a broad range, up to thousands of inner orbit ZLK timescales.
For larger mutual inclinations, secular migration operates over a much broader region of parameter space with $\mathcal{R}\sim 0.05-100$, but most evolutionary pathways become non-secular and potentially unstable---behavior recently identified as an alternative pathway to tidal migration.
Our model predicts hot Jupiters in nearly polar orbits relative to both the host star's stellar equator (stellar obliquities $\sim 60^\circ$–$120^\circ$) and the orbits of the outer two companions. Future Gaia releases and long-term radial velocity campaigns are likely to uncover additional “3+1” systems, providing valuable opportunities to test this migration pathway.
\end{abstract}


\section{Introduction} 

Giant planets are generally expected to form at wide separations from their host stars, beyond the snow line where large planetesimals can coalesce and accrete substantial gaseous envelopes \citep[e.g.,][]{Rafikov2011}. Consequently, close-in giant planets—such as hot and warm Jupiters—are particularly intriguing, as at least some must have migrated inward from their original formation sites. Two main pathways have been proposed to explain their origin: migration through interactions with the protoplanetary disk, and high-eccentricity tidal migration  \citep{dawson2018origins}. Among these, high-eccentricity tidal migration has emerged as a compelling and natural mechanism for producing hot Jupiters \citep[e.g.][]{Rasio1996, wu2003planet}. In this process, a giant planet’s orbital eccentricity is first excited to large values; during subsequent close periapsis passages, tidal dissipation extracts orbital energy and angular momentum, gradually shrinking the orbit until the planet settles close to its host star.

Soon after the discovery of eccentric giant planets such as 16 Cygni Bb \citep{16Cygni}, wide stellar companions were identified as likely sources of eccentricity excitation through the von Zeipel-Lidov-Kozai (ZLK) mechanism \citep{holman97}. Nearly three decades later, most of the known giant planets with extreme eccentricities ($e>0.9$) are found to reside in wide binary systems with separations of order $10^3$ au. A few of these planets, including HD 80606 b \citep{Naef2001}, TIC 241249530 b \citep{Arvind2024}, HIP 66074/Gaia 3b \citep{Sozzetti2023},  reach periastron distances of $\sim$0.03–0.05 au, placing them on the verge of completing their high-e migration journey. In addition to their high eccentricities, many “Kozai-migrated” planets exhibit large stellar obliquities, offering a natural explanation for the substantial spin–orbit misalignments observed in some hot Jupiter systems \citep{Albrecht2021perpendicular}.

The appeal of ZLK migration lies in its simplicity: the system involves only three bodies governed by secular dynamics with short-range forces, allowing for efficient orbit-averaged integrations and analytical insight. This framework has led to a range of theoretical developments and predictions, including early estimates of the stellar obliquity ($\psi$) distribution \citep{Fabrycky2007ZLK}, the role of binary eccentricity in modulating ZLK cycles and broadening the range of inclinations that yield extreme eccentricities \citep{naoz2011hot,naoz2012formation}, steady-state distributions of orbital elements \citep{Petrovich2015}, the onset of chaotic spin–orbit evolution \citep{Storch2014}, among others.

However, the canonical three-body setup—a central star, a cold Jupiter at $\sim$1–10 au, and a distant stellar companion at $\sim$100–1000 au—likely oversimplifies the architecture of most exoplanet systems hosting cold Jupiters. Both formation models \citep[e.g.,][]{Ida2013,Matsumura2021} and radial velocity surveys \citep{Knutson2014,Zhu2022} indicate that multiple giant planets often form and coexist. Observationally, many cold Jupiters show evidence for additional massive companions \citep{zink2023hot}, and their broad eccentricity distribution \citep[e.g.,][]{Butler2006,Kane2012,Kane2024} points to a history of planet–planet scattering among multiple gas giants \citep[e.g.,][]{Juric2008,Raymond2009}. Such scattering episodes usually leave behind two surviving planets, even in the presence of a wide binary companion \citep{Marzari2022}. Moreover, a large parameter space remains between the cold Jupiter and the companion star where undetected substellar objects could reside in dynamically stable orbits—some of which may soon emerge through long-term radial velocity monitoring (see \cite{yang2025hatp7} for a recent example).

Adding an additional planet or substellar companion to the system reduces the direct applicability of these results. The long-term stability and dynamical evolution of the system depend critically on the relative spacing between its constituent objects \citep{Innanen1997,Mustill2017,Avila-Bravo2025}. In closely packed planetary systems, apsidal precession arising from planet--planet interactions can suppress the secular excitation of eccentricity induced by outer companions \citep[e.g.,][]{batygin2011}. However, instabilities following the dissipation of the gas disk can generate widely separated multi-planet configurations \citep{Lu2025}, even in systems that host a stellar companion \citep{Marzari2005}. Such architectures can be treated as hierarchical systems, which are capable of remaining stable over long timescales. These systems are particularly intriguing because interactions among the bodies can excite both the eccentricity and inclination of the planets \citep{Hamers2015}.


In this work, we extend the secular framework traditionally used to describe the von-Zeipel-Lidov-Kozai mechanism in three-body systems \citep{lidov1962evolution,kozai1962secular} to hierarchical four-body systems. We focus on the regime in which the quadrupole potential felt by the inner orbit effectively precesses, which has been analytically described in \cite{Klein2023quad, Klein2024quad, Klein2025quad}. We anticipate that certain regimes will exhibit chaotic behavior when the secular precession frequencies of the inner and intermediate objects become commensurable \citep{Hamers2017quad,Petrovich2017greatly, Grishin2018secular}, while other regimes are expected to produce non-secular excitations \citep{yang2025hatp7}.

Our goal is to investigate secular migration in four-body systems by characterizing the chaotic dynamics (Section \ref{sec:chaos}), mapping the parameter space available for migration (Section \ref{sec:population}), and predicting the stellar obliquities of planets that migrate through this mechanism (Section \ref{sec:obliquity}). In addition, we discuss the applicability of our results to real systems, compare our findings with non-secular four-body migration in \cite{yang2025hatp7}, and place our results in context by contrasting the obliquity distributions of Jupiters migrating through this mechanism with those produced by this alternative migration scenario (Section \ref{sec:discussion}).


\section{Evolution of a Fiducial System}
\label{sec:config}

\begin{table*}
\centering
\begin{tabular}{lll}
\toprule
Parameter  & Definition & Value\\\midrule
$M_0$, $m_1$, $m_2$, $M_3$ & Mass of the system & $1 M_\odot$, $1M_{\mathrm{Jup}}$, $0.03M_\odot$, $1 M_\odot$ \\
$a_1,a_2,a_3$ & Initial semimajor axes & $3, 30, 303.862$ AU\\
$e_1,e_2, e_3$ & Initial eccentricities & $0.05, 0.05, 0.05$ \\
$I_1,I_2,I_3$ & Initial inclinations & $0^\circ, 0^\circ, 20^\circ$\\
$\omega_1, \omega_2, \omega_3$ & Arguments of periapsis & $0^\circ, 0^\circ, 0^\circ$\\
$\Omega_1, \Omega_2, \Omega_3$ & Longitudes of ascending node & $0^\circ, 0^\circ, 0^\circ$\\
$R_{0},r_1$ & Stellar/planetary radii &$1 R_\odot, 1 R_{\mathrm{Jup}}$\\
$k_{2,0},k_{2,1}$ & Love numbers & $0.028, 0.51$ \\
$t_{V,0},t_{V,1}$ & Viscous timescales &  $50\text{ years}, 0.001 \text{ years}$\\
$t_{S,0},t_{S,1}, $ & Spin periods &  $8$ days, $0.5$ day\\
\bottomrule
\end{tabular}
\caption{Fiducial parameters adopted for secular four-body simulations. 
The masses of the host star, planet, brown dwarf, and companion star are denoted by $M_0$, $m_1$, $m_2$, and $M_3$, respectively. 
The radius, Love number, spin period and viscous dissipation timescale of the host star (planet) are represented by $R_0$ ($r_1$), $k_{2,0}$ ($k_{2,1}$), $t_{s0}$ ($t_{s1}$) and $t_{v0}$ ($t_{v1}$), respectively. Orbital elements of the planet, brown dwarf, and companion star, defined with respect to an inertial frame, are indicated by subscripts 1, 2, and 3, respectively.}
\label{tab:fiducial}
\end{table*}

\begin{figure*}
    \centering
    \includegraphics[width=0.98\linewidth]{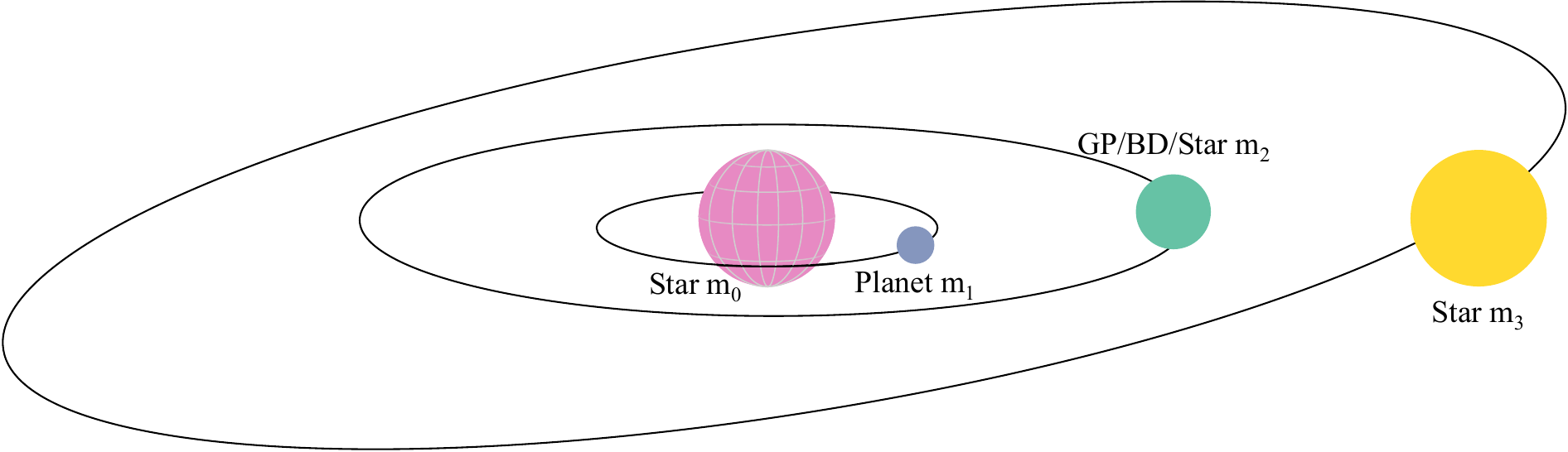}
    \caption{Initial configuration of ``3+1'' hierarchical quadruple systems in the frame of Star 0 (not to scale). The central star is part of a stellar binary and is orbited by the planet of interest, as well as an intermediate perturber, which can be a giant planet, brown dwarf, or low-mass star. As shown in this figure, and throughout most of our analysis, the planet and the intermediate perturber are coplanar, and their common orbital plane has a low mutual inclination with respect to be stellar binary plane.}
    \label{fig:schematic}
\end{figure*}

We consider ``3+1'' quadruple systems, as illustrated in Figure~\ref{fig:schematic}, in which two objects orbit the same star within a wide binary. One of these objects is a planet and the other is an intermediate-mass perturber such as a brown dwarf, super-Jovian planet, or low-mass star. For simplicity, we refer to this perturber as a \emph{brown dwarf}. In Section \ref{sec:chaos}, we show that the dynamical behavior of the system is governed by the ratio of secular timescales instead of by the object types and masses we assign, so the following analysis is not limited to brown dwarf intermediate perturbers.

The masses of the host star, planet, brown dwarf, and companion star are denoted by $M_0$, $m_1$, $m_2$, and $M_3$, respectively. The planet's orbit around the host star is labeled with subscript~1. The brown dwarf's orbit around the center of mass of the planet--host star pair is labeled with subscript~2, and the stellar companion's orbit around the center of mass of the inner triple is labeled with subscript~3.

The orbital elements---semi-major axis $a$, eccentricity $e$, argument of pericenter $\omega$, and longitude of ascending node $\Omega$---follow the same subscript notation. We also define the normalized angular momentum and eccentricity vectors of each orbit as $\mathbf{j}_i$ and $\mathbf{e}_i,$ respectively. Additionally, the mutual inclination between two orbits $I_{kl}$ is given by $\cos^{-1}{(\hat{\mathbf{j}}_k.\hat{\mathbf{j}}_l)}$.

In our model, we assume that the system is hierarchical, i.e., $a_1 \ll a_2 \ll a_3$. To study the long-term evolution of the system, we derive a secular Hamiltonian for the four-body system by expanding the interaction potentials up to octupole order in the semi-major axis ratios and then averaging the retained terms over the mean anomalies of the orbits (see \cite{Hamers2015} for a detailed derivation). 

For simplicity, we ignore coupling terms between the stellar companion and the planet. These terms are small in hierarchical systems and can generally be neglected (see the discussion in Section~2.2 of \cite{Hamers2017quad}). We also ignore the small octupole-order perturbations from the brown dwarf on the stellar companion.


In addition to point-mass dynamics, we include apsidal precession of the planetary orbit due to general relativity and tidal deformations. Apsidal and nodal precession arising from the rotational deformation of both the star and the planet are also taken into account. We further model tidal dissipation in the host star and the planet using the equilibrium tide model. See \cite{Eggleton2001} for the corresponding corrections to the equations of motion. 

The tidal parameters—including the radius, Love number, spin period, and viscous dissipation timescale of the host star (planet)—are denoted by $R_0$ ($r_1$), $k_{2,0}$ ($k_{2,1}$), $t_{S,0}$ ($t_{S,1}$), and $t_{V,0}$ ($t_{V,1}$), respectively (see \cite{Petrovich2015} for a definition of the viscous timescale).

Our goal is to investigate the conditions under which the inner planet can migrate inward through secular-chaos induced high-eccentricity tidal migration in low mutual inclination systems. For our fiducial simulation, we assign system parameters as summarized in Table \ref{tab:fiducial}. 

\begin{figure}
    \centering
    \includegraphics[width=0.98\linewidth]{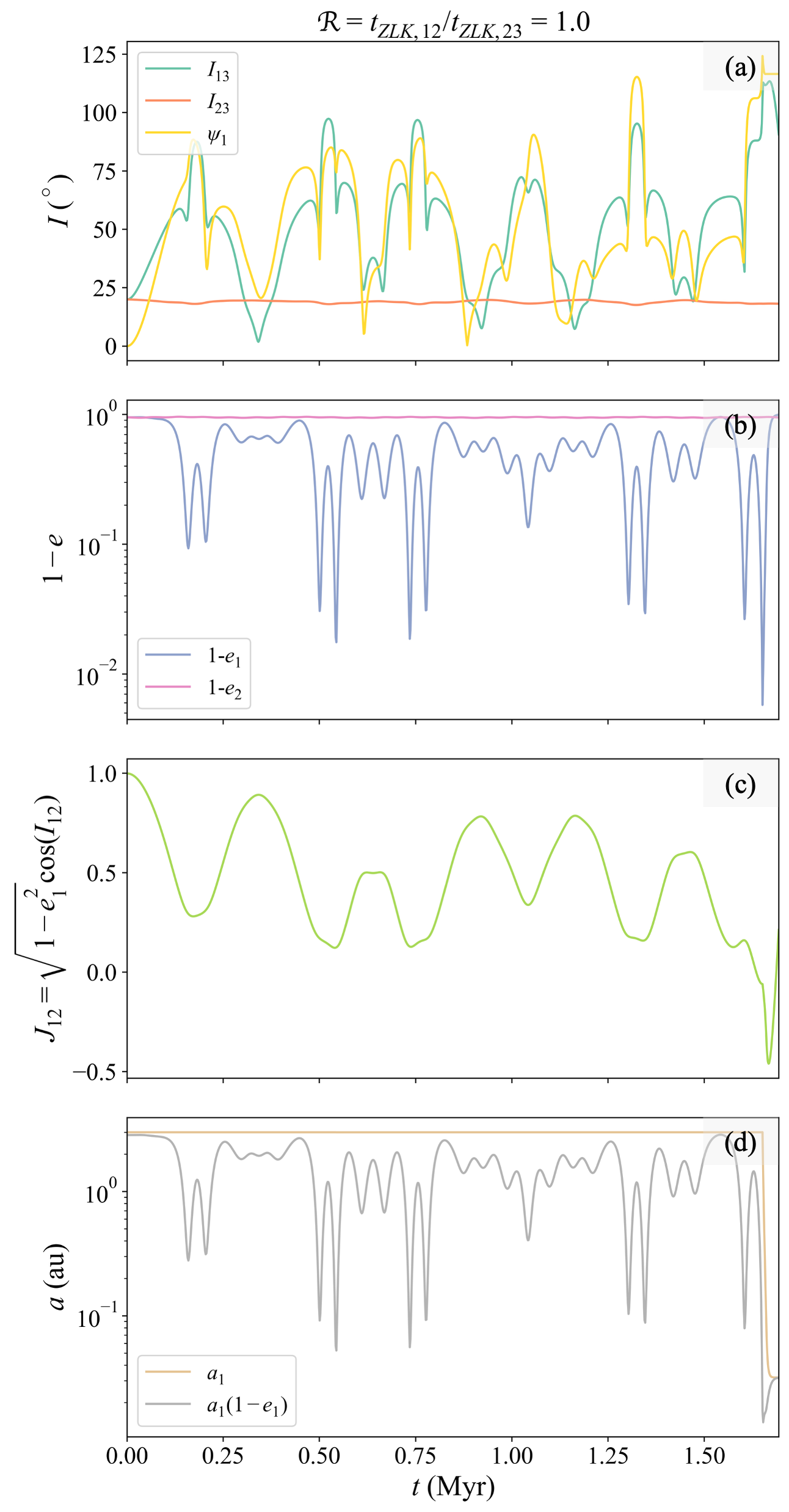}
    \caption{Example of a hot Jupiter forming through high-eccentricity migration triggered by secular chaos. The system parameters are shown in Table \ref{tab:fiducial}. As shown in panels (a) and (b), the inclination and eccentricity of the intermediate brown dwarf perturber undergo small-amplitude precessions, whereas the inclination and eccentricity of the planet are excited simultaneously as $J_{12}=\sqrt{1-e_1^2}\cos I_{12}$ (shown in panel c) is depleted. Eventually, tidal interactions between the star and the planet at close approaches circularize the orbit into a close-in orbit, as evident from the end of panel (d).}
    \label{fig:chaotic_example}
\end{figure}

We integrate our fiducial system using a custom secular code created based on the setup described above, resulting in the evolution shown in Figure~\ref{fig:chaotic_example}. We benchmark our integration against the N-body code TIDYMESS \citep{Boekholt2023tidymess} and find that the two codes produce qualitatively similar evolutionary trajectories for the fiducial system.

From panels~(a) and (b), we see that both the eccentricity and the mutual inclination between the brown dwarf and the stellar companion ($I_{23}$) undergo only small-amplitude oscillations. This behavior is expected because $I_{23} = 20^\circ$ is below the minimum inclination of $\sim 40^\circ$ required to trigger the ZLK mechanism. Meanwhile, the eccentricity of the planet's orbit is excited to extremely high values, approaching unity ($>0.99$). The largest eccentricity spikes occur whenever $I_{12}$ crosses $90^\circ$.

Panel~(c) shows large variations in the component of the normalized angular momentum of the planet along the angular momentum of the brown dwarf, $J_{12}$. As $|J_{12}|$ becomes small, the eccentricity can be excited to near-unity values.

During these high-eccentricity peaks, tidal forces between the star and the planet remove orbital energy, causing the planet’s semi-major axis to shrink (panel~d). Successive episodes of eccentricity excitation therefore lead to successive phases of orbital decay, ultimately producing a close-in, nearly circular orbit with a final semi-major axis of $a_1 \simeq 0.03\,\mathrm{AU}$—successfully forming a hot Jupiter.

From panel~(a), we see that the stellar obliquity of the planet ($\psi_1$) closely follows the inclination of the planet with respect to the stellar companion. As the planet migrates inward, it becomes decoupled from the brown dwarf. Consequently, the obliquity becomes locked at a value just above $115^\circ$.

Overall, our example run shows that the long-term evolution of four body systems in initially near-coplanar systems can be chaotic. In these systems, eccentricity can be excited in the innermost binary, leading to high eccentricity migration of the inner planet. In the following section, we quantify the parameter space where the dynamics are chaotic.

\section{Surfaces of section: chaotic phase-space}
\label{sec:chaos}

We shall study the dynamical behavior of the ``3+1'' quadruple system by means of surfaces of section and explore the emergence of secular chaos. To proceed, we make two simplifying assumptions:
\begin{enumerate}
    \item The orbit of star 3 does not change substantially due to the perturbation from the planet and the brown dwarf, so that all its orbital elements can be considered as static. This is a reasonable assumption because the other objects in the system are much less massive.
    \item The eccentricity of the brown dwarf $e_2$ is consistently small. This is reasonable because, in the low mutual inclination configurations, star 3 only induces small amplitude precessions in the brown dwarf's orbit, as we saw in Section \ref{sec:config}.
\end{enumerate}

Under these assumptions, we move to the frame rotating with 
$\mathbf{j_2}$, the angular momentum vector of the brown dwarf.
By writing the potential in this frame, we can reduce the degrees of freedom of the Hamiltonian to two, similar to the derivation in \cite{Petrovich2017greatly}. 

For small $e_2$, the equation of motion of $\mathbf{j_2}$ can be written as \citep{Liu2015suppression}:
\begin{equation}
    \frac{d\mathbf{j_2}}{dt}=\frac{4}{3t_{\mathrm{ZLK},23}}(\mathbf{j_2}\cdot\mathbf{\hat{j}_3})\mathbf{j_2}\times\mathbf{\hat{j}_3},
\end{equation}
where $t_{\mathrm{ZLK},23}$ is the ZLK timescale given by \citep[e.g.][]{naoz2016eccentric}:
\begin{equation}
    t_{\mathrm{ZLK},ij} = \frac{1}{n_i}\left(\frac{m_0+m_i}{m_j}\right)\left(\frac{a_j}{a_i}\right)^3\left(1-e_j^2\right)^{3/2}
\end{equation}
In other words, the angular momentum vector $\mathbf{j_2}$ is simply precessing around $\mathbf{\hat{j}_3}$  at a constant rate of $4(\mathbf{j_2}\cdot\mathbf{\hat{j}_3})/{3 t_{\mathrm{ZLK},23}}$.
Thus, we can write the dimensionless potential in this frame as
\begin{align}
    \frac{\mathbf{\bar{\Phi}_{rot}}}{\phi_0} 
    &= \frac{1}{6} - e_1^2 - \frac{1}{2}(\mathbf{j_1} \cdot \mathbf{\hat{j}_2})^2+\frac{5}{2}(\mathbf{e}_1 \cdot \mathbf{\hat{j}_2})^2 \notag\\
    &+\frac{4}{3t_{\mathrm{ZLK},23}}\frac{\sqrt{G(m_0+m_1)a_1}}{\phi_0}(\mathbf{\hat{j}_2}\cdot\mathbf{\hat{j}_3})(\mathbf{j_1}\cdot\mathbf{\hat{j}_3}), 
\end{align}
where
\begin{equation}
    \phi_0 = \frac{3GM_2a_1^2}{4a_2^3(1-e_2^2)^{3/2}}
\end{equation}
and the multiplier $(\mathbf{\hat{j}_2}\cdot\mathbf{\hat{j}_3})(\mathbf{j_1}\cdot\mathbf{\hat{j}_3})$ simply becomes
\begin{align}
    \mathcal{R}&\equiv\frac{t_{\mathrm{ZLK},12}}{t_{\mathrm{ZLK},23}} \notag\\
    &=\frac{\sqrt{m_0+m_1}}{\sqrt{m_0+m_1+m_2}}\frac{m_3}{m_2}a_1^{-\frac{3}{2}}a_2^{\frac{9}{2}}a_3^{-3}\frac{(1-e_2^2)^{\frac{3}{2}}}{(1-e_3^2)^{\frac{3}{2}}} \notag\\
    &\simeq\frac{m_3}{m_2}a_1^{-\frac{3}{2}}a_2^{\frac{9}{2}}a_3^{-3}\frac{(1-e_2^2)^{\frac{3}{2}}}{(1-e_3^2)^{\frac{3}{2}}} \\
    &=\frac{m_3}{m_2}\left(\frac{a_2}{a_1}\right)^{\frac{3}{2}}\left(\frac{a_2}{a_3}\right)^{3}\frac{(1-e_2^2)^{\frac{3}{2}}}{(1-e_3^2)^{\frac{3}{2}}}. 
    \label{eq:R}
\end{align}
which was similarly defined in \cite{Hamers2015} and \cite{Grishin2018secular}.


To compute surfaces of section, we move to the rotating reference frame of the brown dwarf. Using the argument of pericenter, $\omega_{12}$, and the longitude of the ascending node, $\Omega_{12}$, of the planetary orbit, as well as the longitude of the ascending node of the stellar orbit, $\Omega_{23}$, all defined with respect to the brown dwarf. 
\footnote{
The unit vector pointing toward the ascending node of the planetary orbit is given by $\hat{\mathbf{n}}_{12}=\mathbf{\hat{j}_2}\times\hat{\mathbf{j}}_1$. The longitude of ascending node and argument of pericenter of the planetary orbit, defined with respect to the rotating reference, can then be written as $\Omega_{12}=\tan^{-1}({\hat{\mathbf{n}}_{12}\cdot(\mathbf{\hat{j}_2}\times\mathbf{\hat{e}_2})/\hat{\mathbf{n}}_{12}\cdot\mathbf{\hat{e}_2}})$ and $\omega_{12}=\cos^{-1}(\hat{\mathbf{e}}_{1}.\hat{\mathbf{n}}_{12})$ respectively. The longitude of ascending node of the stellar orbit is also similarly defined: $\Omega_{32}=\tan^{-1}({\hat{\mathbf{n}}_{32}\cdot(\mathbf{\hat{j}_2}\times\mathbf{\hat{e}_2})/\hat{\mathbf{n}}_{32}\cdot\mathbf{\hat{e}_2}})$, where $\hat{\mathbf{n}}_{32}=\mathbf{\hat{j}_2}\times\hat{\mathbf{j}}_3$ is the unit vector pointing toward the direction of longitude of the ascending node of the stellar orbit.}, the dimensionless Hamiltonian can be written as:


\begin{align}
    \tilde{H}_{\mathrm{tot}} &= \tilde{H}_{\mathrm{ZLK},12} + \mathcal{R}\tilde{H}_{\mathrm{rot}} \notag \\
    &=-\frac{1}{3}-\frac{1}{2}e_1^2+(\frac{1}{2}+2e_1^2-\frac{5}{2}e_1^2\cos^2{\omega_{12}})\sin^2{I_{12}}\notag \\
    &+\mathcal{R}(1-e_1^2)^{1/2}[\cos{I_{12}}\cos^2{I_{32}} \notag \\
    &+\frac{1}{2}\sin{I_{12}}\sin{2I_{32}}\cos{(\Omega_{12}-\Omega_{32})}]
    \label{eq:Htot},
\end{align}
where we identify the first term as the ZLK Hamiltonian, and the second term is the correction from the rotating frame induced by orbit 3. The parameter $\mathcal{R}$ controls the relative importance of the second component and will be used throughout the rest of the paper.

For simplicity, we integrate the vectorial equations of motion in the rotating frame given by \citep{Petrovich2017greatly}:
\begin{eqnarray}
\frac{d\mathbf{e_1}}{d\tau} &=& 2\mathbf{j_1}\cdot\mathbf{e_1} -5(\mathbf{e}_1\cdot\mathbf{\hat{j}_2})\mathbf{j_1}\times\mathbf{\hat{j}_2} + (\mathbf{j_1}\cdot\mathbf{\hat{j}_2})\mathbf{e}_1\times\mathbf{\hat{j}_2} \\ \notag
    &&-\mathcal{R}(\mathbf{\hat{j}_3}\cdot\mathbf{\hat{j}_2})\mathbf{e_1}\times\mathbf{\hat{j}_3},\\
    \label{eq:edot}
        \frac{d\mathbf{j_1}}{d\tau} &=& (\mathbf{j_1}\cdot\mathbf{\hat{j}_2})\mathbf{j_1}\times\mathbf{\hat{j}_2}-5(\mathbf{e_1}\cdot\mathbf{\hat{j}_2})\mathbf{e_1}\times\mathbf{\hat{j}_2} \\ \notag
    &&-\mathcal{R}(\mathbf{\hat{j}_3}\cdot\mathbf{\hat{j}_2})\mathbf{j_1}\times\mathbf{\hat{j}_3},
    \label{eq:jdot}
\end{eqnarray}
with $\tau=t/t_{\mathrm{ZLK},12}$ and create surfaces of section to analyze the dynamical behavior. 

\begin{figure*}
    \centering
    \includegraphics[width=\textwidth]{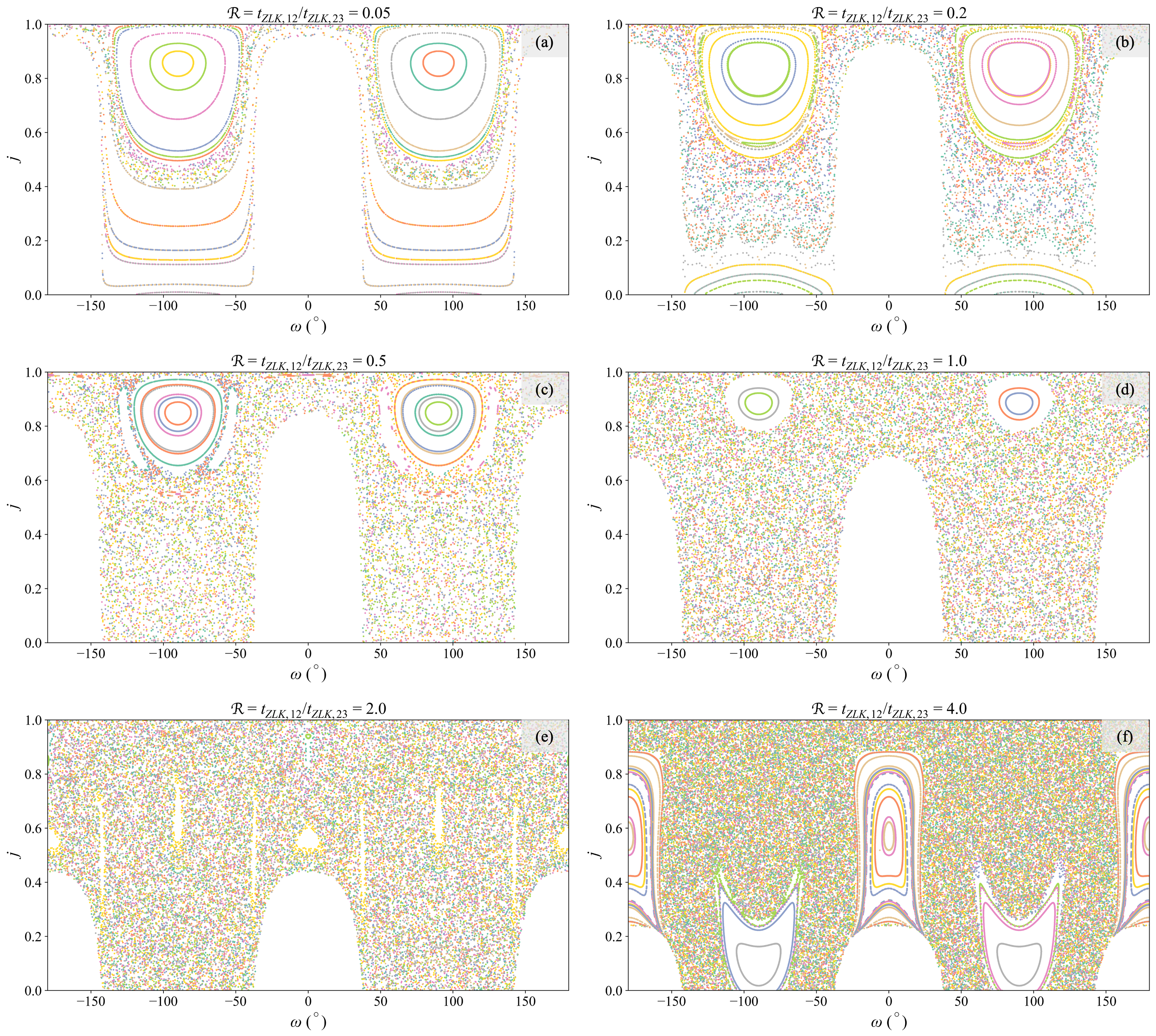}
    \caption{Surface of sections in $\omega$-j space of the rotating frame Hamiltonian in Equation \ref{eq:Htot}. These figures are created by integrating equations of motion in Equations y-z. The Hamiltonian and the system parameters are fixed as $H=0.1$, $I_{23}=20^\circ$, $\Omega_{23}=0^\circ$. The completely empty regions in the figures are the prohibited region by conservation of H. When $\mathcal{R}=0.05$, there is only a small band of chaos; most trajectories resemble the ZLK Hamiltonian trajectories. As $\mathcal{R}$ increases, the chaotic region increases, until when $\mathcal{R}=1$, there is only a small librating region, the rest of the allowed region is chaotic.}
    \label{fig:sections}
\end{figure*}

Our procedure to obtain the surfaces of section in Figure \ref{fig:sections} is as follows.
We keep the constants $\tilde{H}_{\mathrm{tot}}$, $I_{23}$, and $\Omega_{23}$ fixed across all panels, and vary $\mathcal{R}$ from panel to panel to observe how it affects the behavior. We create each panel by initializing a batch of runs by sampling initial $j_1=(1-e_1^2)^{1/2}$ and $\omega_{12}$. For brevity, we drop the subscripts for $j_1 (\equiv j)$ and $\omega_{12}(\equiv\omega)$.  We solve for the initial mutual inclination $I_{12}$ using Equation (\ref{eq:Htot}). We then integrate the equations of motion (\ref{eq:edot}) and (\ref{eq:jdot}), and record a instance when $\Omega_{12}$ crosses 0 in the direction where $d\Omega_{12}/dt > 0$. These points make up trajectories in $j-\omega$ space. 

We show an example set of surfaces of sections in Figure \ref{fig:sections}, keeping $\tilde{H}_{\mathrm{tot}}=0.10$. We can summarize the behavior as follows:
\begin{enumerate}
    \item In the $\mathcal{R} \ll 1$ regime (panel a), $\tilde{H}_{\mathrm{tot}}$ is dominated by $\tilde{H}_{\mathrm{ZLK}}$, and the trajectories resemble those of the ZLK surfaces of section with librating regions at
    $\omega =\pm90^\circ$. The perturbation from $\tilde{H}_{\mathrm{rot}}$ creates chaotic bands near the separatrix, which grow larger as $\mathcal{R}$ increases. It also creates another libration island at very high eccentricities, more readily observable in panel (b). Due to the island, the diffusion from $j\sim1$ to $j\sim0$ would be inhibited. 
    \item When $\mathcal{R} \sim 1$ (panels c and d), the libration island at $j\sim 0$ disappears, and diffusion from $j\sim 1$ to $j\sim 0$ is allowed for all trajectories. In fact, for $\mathcal{R}=1$ (panel d), nearly all the phase-space is chaotic, and diffusion that allows for tidal migration is guaranteed.

    \item When $\mathcal{R} > 1$, the effect of $\tilde{H}_{\mathrm{rot}}$ begins to dominate. Narrow librating regions due to higher order resonances began to form near $\omega =0^\circ$, $\pm45^\circ$, $\pm90^\circ$, $\pm135^\circ$, and $\pm180^\circ$ (panel e). As $\mathcal{R}$ increases further, these librating regions widen (panel f).
\end{enumerate} 

In summary, the precession induced by orbit 3 gives rise to higher-order resonances in the ZLK phase-space portraits, allowing for imminent chaotic diffusion from circular to near parabolic orbits when  $\mathcal{R}\sim 1$.

\section{Population Statistics}
\label{sec:population}
While Section \ref{sec:config} presented a representative example of secular chaos producing a hot Jupiter in a “3+1” architecture, we now assess the effectiveness and efficiency of this mechanism at a population level in more general configurations. The behavior of secular chaos is determined by the value of $\mathcal{R}$, as described in Section \ref{sec:chaos}, rather than by the specific values of the system parameter. Accordingly, for a population-level statistical analysis, we sample over $\mathcal{R}$ space instead of exploring the whole space of physical parameters. In real systems, this simplification is complicated by the inclusion of short-range forces such as tidal dissipation. See Section \ref{sec:degeneracy} for a discussion. Here, for simplicity, we consider purely four-body secular dynamics and neglect tidal forces and relativistic precession, making the evolution scale-invariant. We use the maximum eccentricity attained by the planet as a proxy for its ability to undergo high-eccentricity migration and eventually form a hot Jupiter.

In \S\ref{sec:timetounity}, we evaluate the efficiency of the mechanism by measuring the time required for the planet’s eccentricity to reach near-unity. In \S\ref{sec:vary_Imut}, we examine how increasing the mutual inclination between the brown dwarf and the stellar companion expands—or restricts—the region of parameter space in which the planet can be driven to near-parabolic eccentricities.

\subsection{Time to excite planet eccentricity to near-unity}
\label{sec:timetounity}
We want to investigate the efficiency of secular chaos in ``3+1'' systems by examining the time to excite the planet's eccentricity to near-unity, which we define as $e_1=1-10^{-2}=0.99$. Specifically, we want to see how this efficiency varies with different values of $\mathcal{R}$. To explore this, we set up a batch of simulations for each value of $\mathcal{R} = \{0.5, 0.75, 1.0, 2.0\}$. Each batch of runs contains 200 simulations. We set $I_1=I_2=20^\circ$ and $I_3=0^\circ$. Then we randomize $\omega_1$ and $\Omega_1$ from $(-180^\circ, 180^\circ)$. $I_{23}$ is fixed at $20^\circ$. As $\Omega_1$ is varied, $I_{12}$ can range from $0^\circ$ to $40^\circ$, staying below the ZLK limit. Other system parameters are taken from Table \ref{tab:fiducial}, except for $a_3$, which is calculated based on the value of $\mathcal{R}$.

\begin{figure}
    \centering    
    \includegraphics[width=0.98\linewidth]{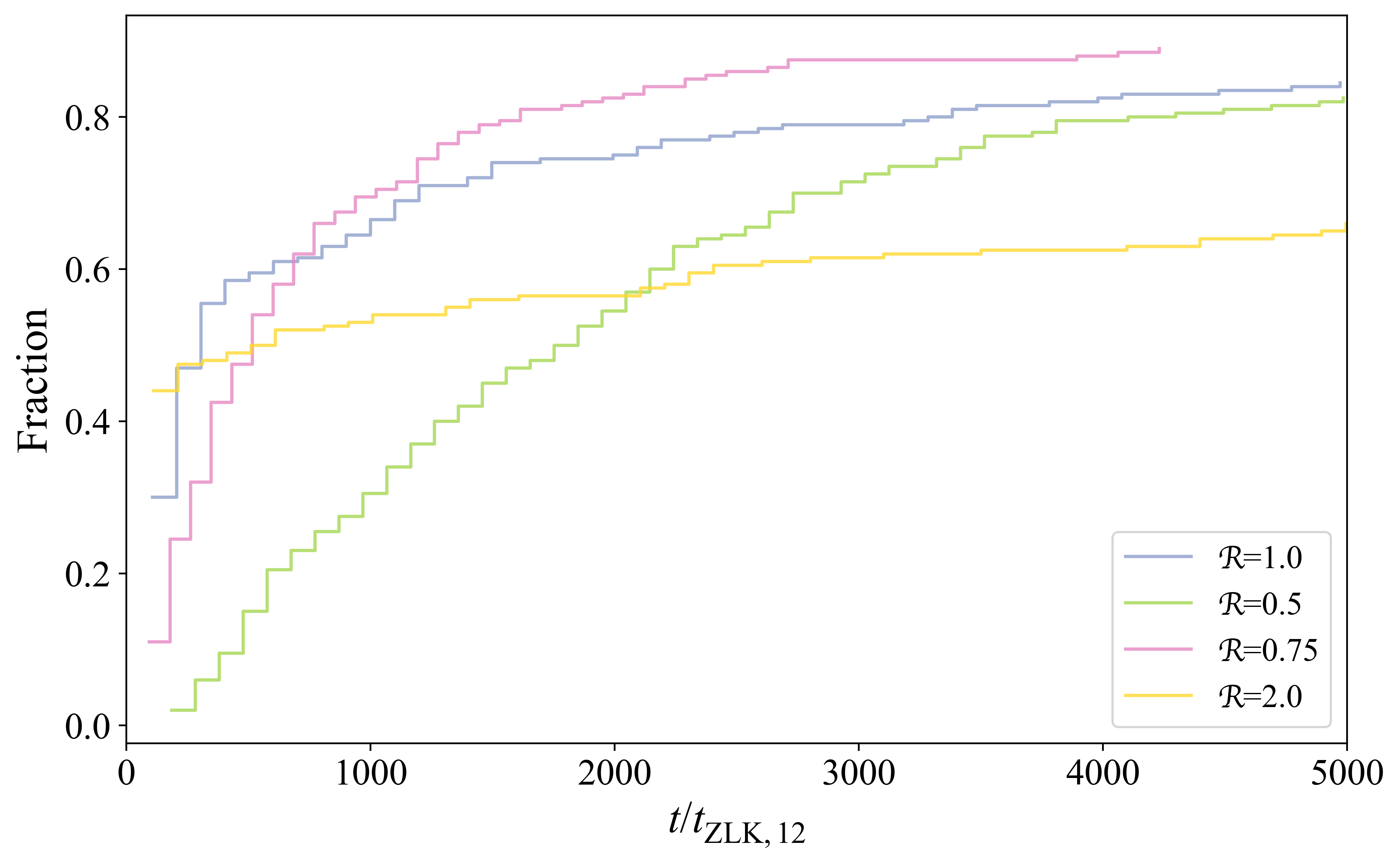}
    \caption{The fraction of simulations where the planet's eccentricity reaches $0.99$ from an initial eccentricity of $0.05$ as a function of normalized time. We set $I_1=I_2=20^\circ$, $I_3=0^\circ$, and randomly sample $\omega_1$, $\Omega_1$. All other system parameters are shown in Table \ref{tab:fiducial}, except for $a_3$, which is calculated based on the value of $\mathcal{R}$. All simulations have a time limit of $5000t_{\mathrm{ZLK},12}$. While the time it takes for planets to be excited to near-unity eccentricity decreases with increasing $\mathcal{R}$, the fraction of successfully excited systems doesn't follow a clear monotonic trend.}
    \label{fig:timetounity}
\end{figure}

A run is considered to be successful if the planet's eccentricity reaches $e_1 = 0.99$ within a time limit of $5000t_{\mathrm{ZLK},12}$. We show the fraction of simulations successfully reaching $e_1=0.99$ in Figure \ref{fig:timetounity}, where each color corresponds to a different batch with a certain value of $\mathcal{R}$. A total of 200 simulations were run for each batch. The x-axis is time normalized by $t_{\mathrm{ZLK},12}$, which is fixed at a value of $\sim2.8\times10^4$ years for all runs. 

As $\mathcal{R}$ increases, the time it takes to excite the planet's eccentricity drops significantly. This is expected because as $\mathcal{R} = \frac{t_{\mathrm{ZLK},12}}{t_{\mathrm{ZLK},23}}$ increases, $t_{\mathrm{ZLK},23}$ decreases, and thus the timescale of the system is shorter. In Figure \ref{fig:timetounity}, we can see this same effect in the first bin. However, the fraction of systems that can be excited to $e_1=0.99$ does not follow a simple monotonic trend, so this clear trend is lost for the rest of the figure. 

\begin{figure}
    \centering
    \includegraphics[width=0.98\linewidth]{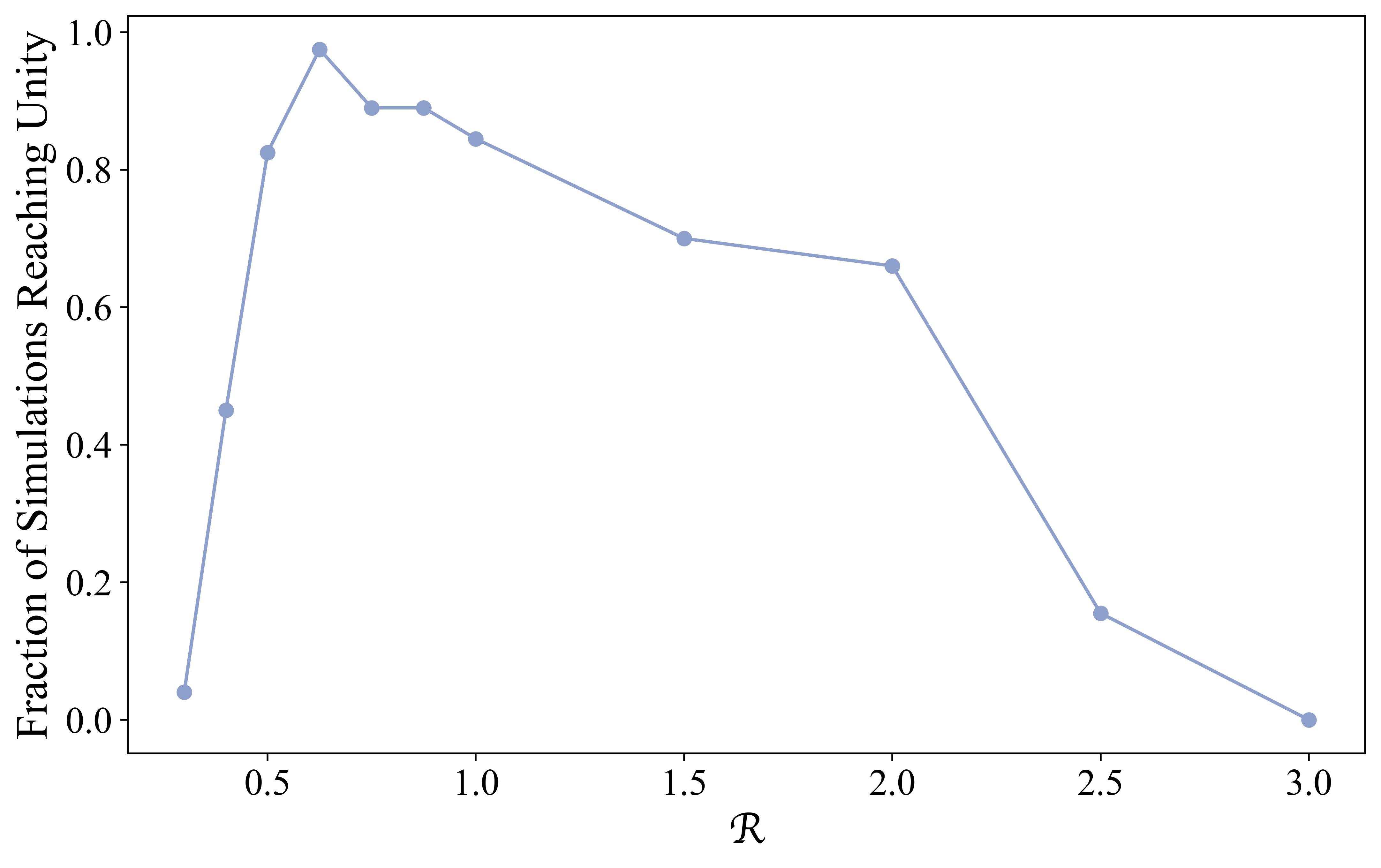}
    \caption{The fraction of simulations in which the planet reaches near-unity eccentricity ($e_1>1-10^{-2}$) as a function of $\mathcal{R}$. The fraction increases with $\mathcal{R}$ in the low $\mathcal{R}$ regime, plateaus around $\mathcal{R}=0.6-1.0$, and decreases with $\mathcal{R}$ in the high $\mathcal{R}$ regime.}
    \label{fig:fraction}
\end{figure}

To better show how the fraction of successful systems is affected by $\mathcal{R}$, we create an additional seven batches of runs, with $\mathcal{R} = \{0.3, 0.4, 0.625, 0.875, 1.5, 2.5, 3.0\}$, to better sample $\mathcal{R}$ space. For each batch, we calculate the fraction of successful runs. The results are shown in Figure \ref{fig:fraction}. The fraction first increases with $\mathcal{R}$, plateaus at around $\mathcal{R} = 0.6-1.0$, and then decreases with $\mathcal{R}$. The increasing trend at small $\mathcal{R}$ can be explained by the dynamical behavior analysis in Section \ref{sec:chaos}. When $\mathcal{R}$ is very small, the ZLK Hamiltonian dominates, and the planet simply undergoes apsidal and nodal precessions without significant eccentricity excitation in the low mutual inclination case we are looking at here ($I_{12} < 40^\circ$). Chaotic behavior arises as $\mathcal{R}$ increases, and the planet's eccentricity can be excited to near-unity. The decreasing trend at large $\mathcal{R}$ can be explained as such: when $t_{\mathrm{ZLK},23}$ is much larger than $t_{\mathrm{ZLK},12}$, from the perspective of the planet, $\mathbf{j_2}$ is precessing so fast that it is approximately effectively pointing in the direction of $\mathbf{n_3}$. Therefore, the planet is subject to a ZLK Hamiltonian in which the mutual inclination is $I_{13} = 20^\circ$. This is below the critical ZLK limit of $40^\circ$, and the planet's eccentricity cannot be excited to near-unity.

\subsection{Varying mutual inclination}
\label{sec:vary_Imut}

\begin{figure}
    \centering
    \includegraphics[width=0.98\linewidth]{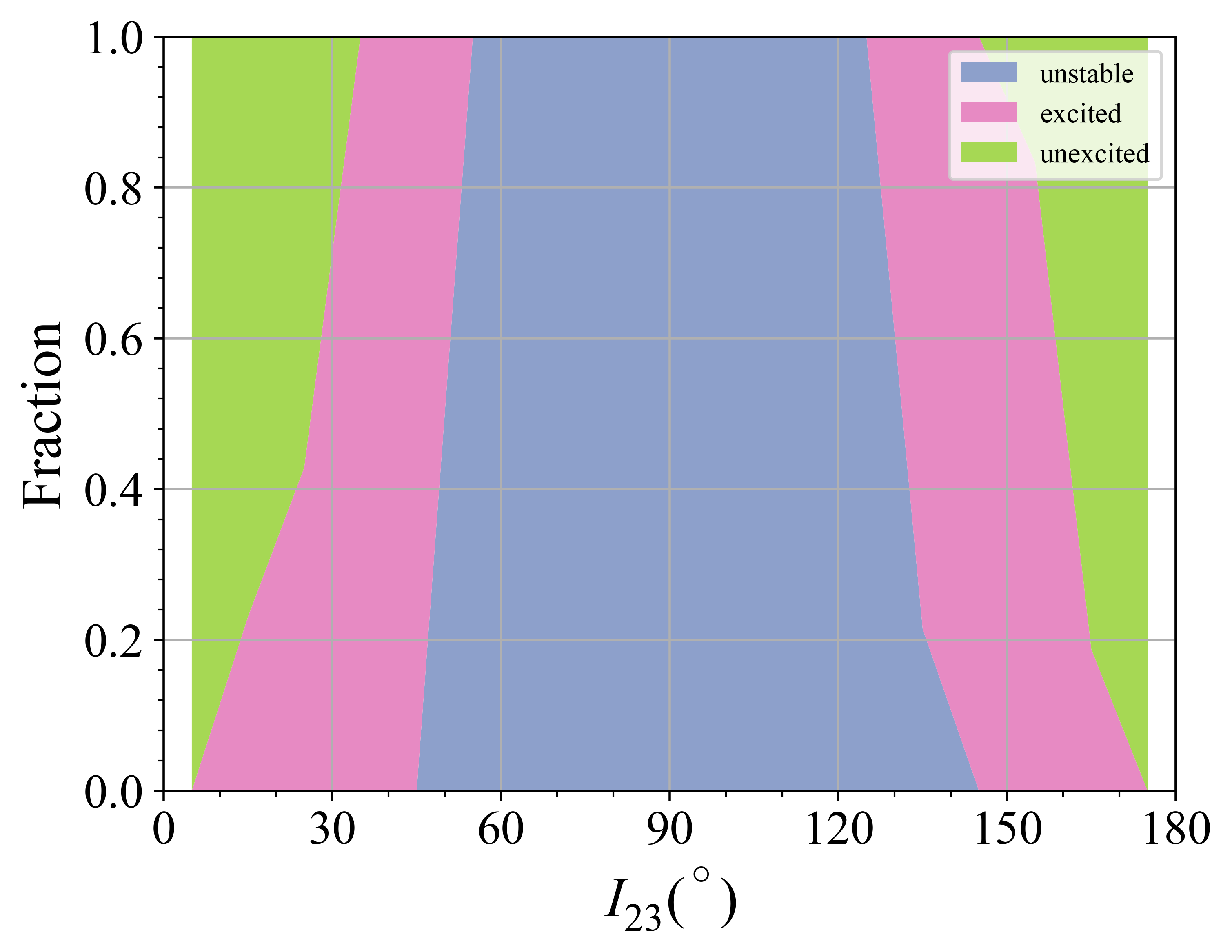}
    \caption{Fractional composition of outcomes of simulations as a function of initial $I_{23}$. We randomize initial $I_{23}$ ($I_{12}$ is kept at $0^\circ$), $\omega_1$, $\Omega_1$ and table all other parameters from Table \ref{tab:fiducial} except for $a_2$ (set to $50$ AU) and $a_3$ (calculated from $\mathcal{R} = 1$). As initial $I_{23}$ approaches $90^\circ$, planets are more likely to be excited to near-unity eccentricity, but systems are potentially go unstable when $I_{23}$ is above the ZLK threshold.}
    \label{fig:fractioninc}
\end{figure}

\begin{figure}
    \centering
    \includegraphics[width=0.98\linewidth]{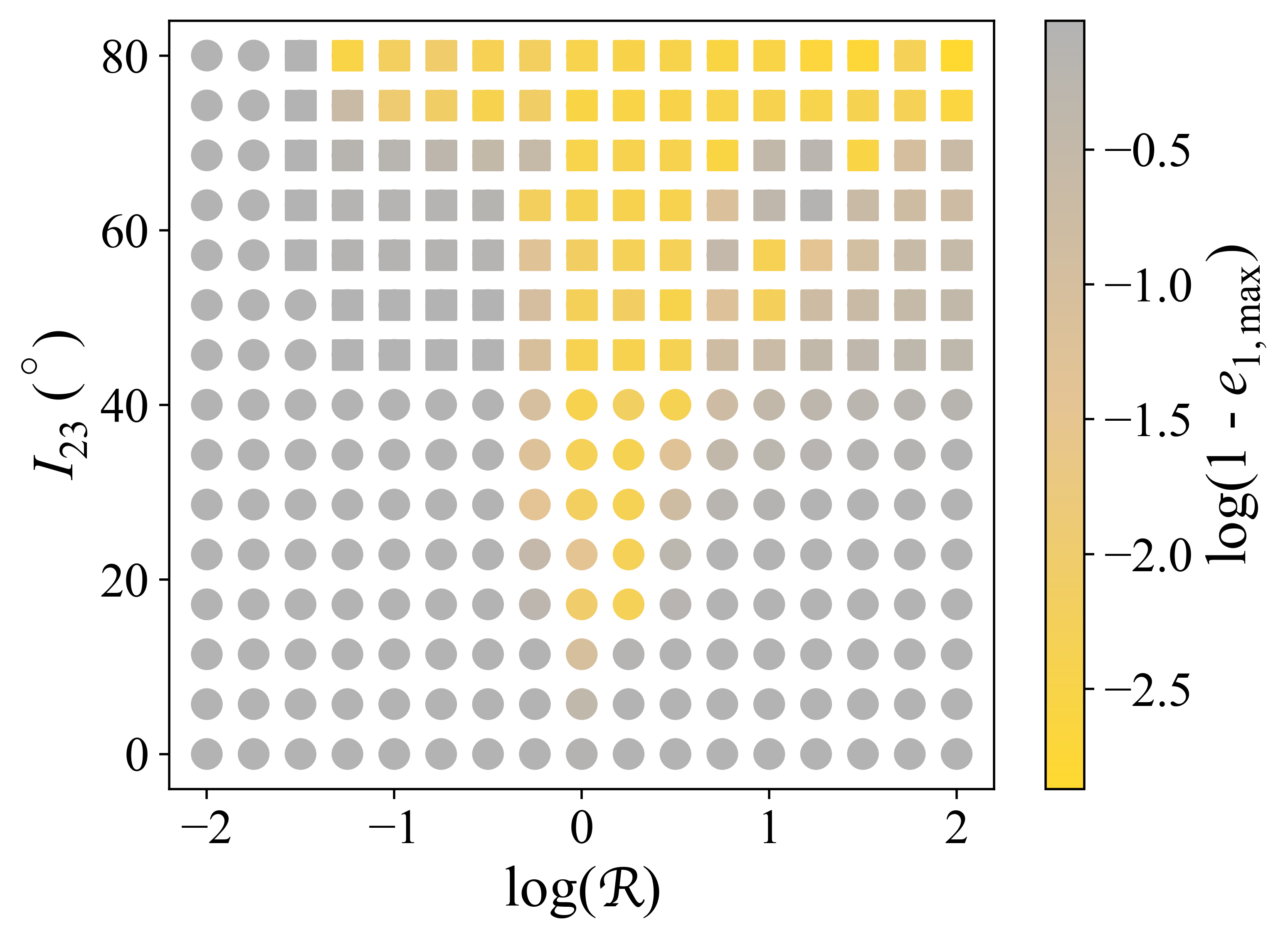}
    \caption{The $\mathcal{R}$ parameter space in which planet eccentricity can be excited with varying initial $I_{23}$. The color bar shows the extent to which eccentricity is excited. The square dots indicate runs that are unstable according to the \cite{Tory2022instability} criteria for instability. All system parameters are taken from Table \ref{tab:fiducial} except for initial $I_{23}$ ($I_{12}$ is kept at $0^\circ$), $a_2$, and $a_3$. $a_2$ is set to $50$ AU. $a_3$ is calculated based on the value of $\mathcal{R}$. As $I_{23}$ increases, the intermediate perturber is not just undergoing small amplitude precessions, but instead ZLK oscillations. The system becomes chaotic for values of $\mathcal{R}$ that are not close to 1.}
    \label{fig:increaseinc}
\end{figure}

Next, we explore how lifting the $I_{23}=20^\circ$ condition affects the effectiveness of secular chaos in ``3+1'' systems. When $I_{23}$ is large, ZLK oscillations can cause the brown dwarf's eccentricity and inclination to be excited, and more chaotic/unstable behavior is expected to arise \citep{Hadden2018,Bhaskar2024}. For this reason, we check for instability between the planet and the brown dwarf at every timestep throughout this section with the criterion from \cite{Tory2022instability}. The \cite{Tory2022instability} criterion is selected because it works for a wide range of mass ratios. To reduce the effect of instabilities, we change $a_2$ to $50$ AU. As shown in Figure \ref{fig:fractioninc} and \ref{fig:increaseinc}, some of our runs have been flagged at least once as unstable within the time limit of $50 t_{\mathrm{ZLK},12}$. Note that systems flagged as unstable by the instability criterion are not necessarily immediately unstable. The criterion merely indicates that the system is susceptible to future instability. Some systems may evolve back into a stable configuration, while others may undergo tidal circularization and form a hot Jupiter before instability actually occurs.

Firstly, keeping $\mathcal{R}$ fixed at 1 and initial $I_{12}$ fixed at $0^\circ$, we randomly sample $I_{23}$ uniformly from $0^\circ$ to $180^\circ$ to lift the $I_{23}=20^\circ$ condition. For each sampling, we also randomly sample initial $\omega_1$ and $\Omega_1$ from (-$180^\circ$, $180^\circ$). All other parameters are fixed at the values in Table \ref{tab:fiducial} except for $a_2$ and $a_3$, which are now 50 AU and 654 AU, respectively. We run 200 simulations with a time limit of 100 $t_{\mathrm{ZLK},12}$ and plot the fractional composition of the outcome as a function of initial $I_{23}$ in Figure \ref{fig:fractioninc}. The three outcomes are unexcited, in which the eccentricity of the planet is never excited to $0.99$, excited, in which the eccentricity of the planet is excited to $0.99$, and unstable, in which the system is flagged as potentially unstable at least once. 

We can see from Figure \ref{fig:fractioninc} that initial $I_{23}$ closer to $90^\circ$ increases the likelihood that the planet can be excited to near-unity eccentricity, but when initial $I_{23}$ is above the ZLK threshold ($40^\circ<I_{23}<160^\circ$), more systems can become potentially unstable as $I_{23}$ approaches $90^\circ$. 

We also examine how relaxing the $I_{23}=20^\circ$ condition affects the system's behavior when $\mathcal{R}$ spans a wide range of values. To accomplish this, we vary $I_{23}$ from $0^\circ$ to $80^\circ$ uniformly over 15 points and vary $\mathcal{R}$ from $10^{-2}$ to $10^2$ uniformly in log space over 17 points. The system parameters of all runs are taken from Table \ref{tab:fiducial} except for $I_{23}$, $a_2$, and $a_3$. $a_2$ is set to 50 AU and $a_3$ is calculated based on the value of $\mathcal{R}$. We continue to use the maximum eccentricity reached by the planet as the indicator of dynamical behavior. 

The results of this set of simulations are shown in Figure \ref{fig:increaseinc}, where the x and y-axis show the 2D grid we sample over, and the color reflects the maximum eccentricity reached by the planet. At around $I_{23}\sim20^\circ$, we observe the behavior shown in Section \ref{sec:timetounity}, the system is most capable of exciting $e_1$ to near-unity around $\mathcal{R} \sim 1$. The system switches to the non-chaotic regime as $\mathcal{R}$ is increased or decreased. However, as $I_{23}$ is increased, the $\mathcal{R}$ space in which the system is chaotic widens significantly. This is expected because larger amplitude precessions of the brown dwarf, as well as eccentricity and inclination variations, cause the system to be more chaotic and less well-described by our analysis in Section \ref{sec:chaos}. In real systems, $I_{23}$ can span a wide range of values, so exciting planetary eccentricity through secular chaos is not limited to the small parameter space of $\mathcal{R}\sim1$. However, when $I_{23}>40^\circ$, the system can potentially become unstable.

\section{Predicted Stellar Obliquities and Inclinations}
\label{sec:obliquity}

Once the planet’s orbit circularizes and shrinks, its final orbital inclination and stellar obliquity ($\psi$) become effectively “fossilized,” providing a powerful diagnostic for distinguishing among tidal migration pathways. In this section, we perform an ensemble analysis to predict the obliquity distribution that emerges from secular chaos in “3+1” systems.

To investigate this, we continue using our secular four-body integration code but now include tidal forces and GR, allowing planets to complete high-eccentricity migration. For each value of $\mathcal{R}=\{0.5,0.7,0.9,1.0\}$, we run a batch of 400 simulations. We maintain a low-mutual-inclination configuration by fixing $I_{12}\approx 0^\circ$—implemented by setting $I_1=I_2=0.01^\circ$—and drawing $I_3$ isotropically between $0^\circ$ and $39.2^\circ$. We randomize $\omega_1$ and $\Omega_1$ uniformly from $-180^\circ$ to $180^\circ$. All other orbital parameters follow Table \ref{tab:fiducial}, except for $a_3$, which we compute directly from the chosen value of $\mathcal{R}$. 

For comparison, we also generate a standard ZLK distribution using the parameters in Table \ref{tab:fiducial}, but removing star 3. In these runs, we set $I_2=0^\circ$, randomize $\omega_1$ and $\Omega_1$, and draw $I_1$ isotropically above the ZLK critical angle of $39.2^\circ$, performing 1800 simulations. All integrations are run up to a maximum of $5000\,t_{\mathrm{ZLK},12}$, and we classify a planet as a hot Jupiter once it satisfies $a_1<0.10$ au and $e_1<0.05$.

\begin{figure}
    \centering
    \includegraphics[width=0.98\linewidth]{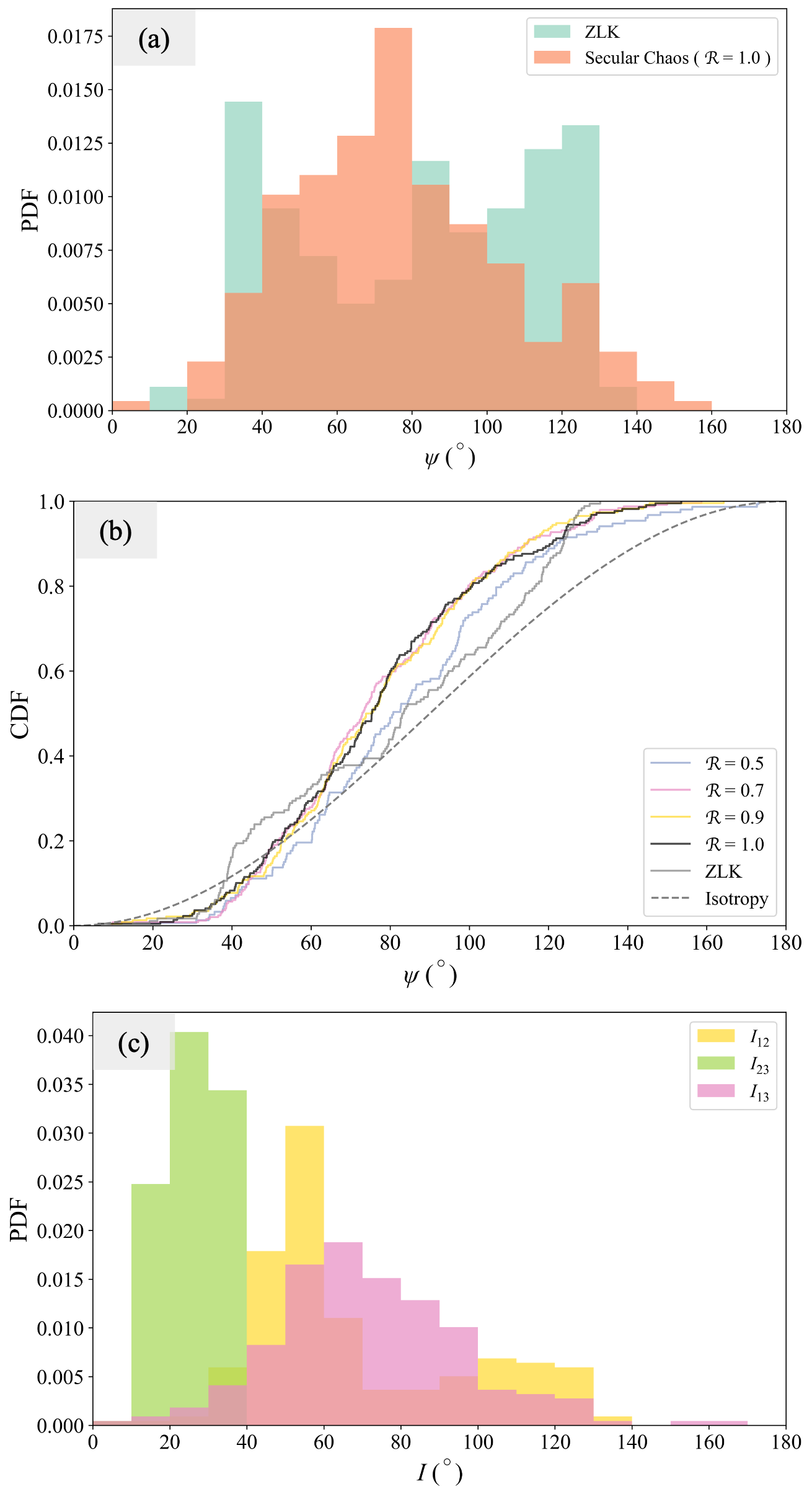}
    \caption{The distribution of final stellar obliquity obtained by the planet and mutual inclinations in the system after the planet evolves into a hot Jupiter. We set $I_1 = I_2 \approx 0^\circ$ and isotropically sample $I_3$ below $39.2^\circ$. $\omega_1$ and $\Omega_1$ are randomly sampled.
    All other system parameters are taken from Table \ref{tab:fiducial} except for $a_3$, which is calculated based on the value of $\mathcal{R}$. All simulations have a time limit of $5000 t_{\mathrm{ZLK},12}$. Panel (a) shows an obliquity distribution with $\mathcal{R} = 1$ and a ZLK obliquity distribution. Panel (b) shows the CDF of two distributions in panel (a), as well as distributions of $\mathcal{R} = \{0.5, 0.7, 0.9\}$ and the theoretical line of isotropy. Panel (c) shows the mutual inclination distributions with $\mathcal{R} = 1$. We can see that the obliquity distribution from secular chaos in ``3+1'' systems is centered around a little less than $90^\circ$, and this result is insensitive to the value of $\mathcal{R}$. This distribution is qualitatively different from the ZLK distribution and isotropy. Similarly, $I_{12}$ and $I_{13}$ are centered around a little less than $90^\circ$, and the $I_{23}$ distribution reflects the sampling.}
    \label{fig:obliquity}
\end{figure}

We present the final obliquity distributions of hot Jupiters in Figure \ref{fig:obliquity}. Panel (a) compares the classical ZLK distribution with the $\mathcal{R}=1$ secular-chaos distribution, revealing a clear qualitative difference: the quadrupole-level ZLK mechanism produces the expected bimodal obliquity distribution \citep[e.g.,][]{Fabrycky2007ZLK}, whereas secular chaos yields a single-peaked distribution centered near $\sim 80^\circ$ (median). Panel (b) shows the corresponding CDFs for several values of $\mathcal{R}$, demonstrating that all secular-chaos distributions are broadly similar to one another and distinctly different from both the ZLK distribution and the isotropic expectation.

Panel (c) shows the final mutual inclination distributions for the $\mathcal{R}=1$ simulations. The angles $I_{12}$ and $I_{13}$ are centered at $\sim60^\circ$ and $\sim70^\circ$, respectively, slightly below $80^\circ$, mirroring the obliquity distribution, while $I_{23}$ retains the imprint of its initial sampling below the ZLK critical angle.

In summary, secular chaos in “3+1” systems naturally produces near-perpendicular planetary orbits while allowing the two outer bodies to remain on low- to moderate–mutual-inclination configurations.

\section{Discussion}
\label{sec:discussion}

We have demonstrated that secular chaos in ``3+1'' can induce tidal migration for $\mathcal{R}$ is order unity with hot Jupiters achieving near polar orbits. In this section, we discuss possible candidates affected by secular chaos (\S\ref{sec:application}), how the stellar obliquity distributions compare to observations (\S\ref{sec:psi_dist}), and how our model compares to other models in quadruple systems (\S\ref{sec:compYang2025}).

\begin{table*}
\centering
\begin{tabular}{lll}
\toprule
Parameter  & Definition & Value\\\midrule
$M_0$, $m_1$, $m_2$, $M_3$ & Mass of the system & $0.976 M_\odot$, $1.703M_{\mathrm{Jup}}$, $55M_{\mathrm{Jup}}$, $0.556 M_\odot$ \\
$a_1,a_2,a_3$ & Initial semimajor axes & $1.28, 42, 1078.17$ AU\\
$e_1,e_2, e_3$ & Initial eccentricities & $0.05, 0.3, 0.855$ \\
$I_1,I_2,I_3$ & Initial inclinations & $0^\circ, 0^\circ, 15^\circ$\\
$\omega_1, \omega_2, \omega_3$ & Arguments of periapsis & $0^\circ, 0^\circ, 0^\circ$\\
$\Omega_1, \Omega_2, \Omega_3$ & Longitudes of ascending node & $0^\circ, 0^\circ, 0^\circ$\\
$R_{0},r_1$ & Stellar/planetary radii &$1 R_\odot, 1 R_{\mathrm{Jup}}$\\
$k_{2,0},k_{2,1}$ & Love numbers & $0.028, 0.5$ \\
$t_{S,0},t_{S,1}, $ & Spin periods &  $10$ days, $0.5$ day\\
$Q_0, Q_1$ & Tidal Quality Factor & $10^6$, $100$\\
$C_0,C_{\mathrm{1}}$ & Gyroradii &  $0.08, 0.25$\\
\bottomrule
\end{tabular}
\caption{Values used for N-Body simulation of the HD 4113 system \citep{Tamuz2008HD4113, Mugrauer2014HD4113B, Cheetham2018hd4113, Stassun2019hd4113B, Feng2022hd4113}.}
\label{tab:hd4113}
\end{table*}

\subsection{Applicability to real systems}
\label{sec:application}

Just as HD 80606 b has become the poster child for classical ZLK migration, it would be valuable to identify systems whose architectures resemble those expected to produce high eccentricities through secular chaos. This is challenging, however, because current detection techniques are not well suited to finding planets or brown dwarfs located between the cold Jupiter and the wide stellar companion—precisely the region where an intermediate perturber would reside.


\paragraph{HD 4113} is a promising example, consisting of an extremely eccentric cold Jupiter, a brown dwarf, and a wide stellar companion \citep{Tamuz2008HD4113, Mugrauer2014HD4113B, Cheetham2018hd4113, Stassun2019hd4113B, Feng2022hd4113}. HD 4113 A is a G-type star of roughly one solar mass. It hosts a proto–hot Jupiter, HD 4113 Ab, with a minimum mass of about $1.7,M_{\rm Jup}$ on a $\sim$1.3 AU orbit with eccentricity $\simeq 0.9$ \citep{Feng2022hd4113}. The system also contains a brown dwarf, HD 4113 C, whose orbit we constrain with archival astrometry and radial velocity data, see Appendix \ref{sec:hd4113_orbit}. The wide M-dwarf companion, HD 4113 B, has a mass of about $0.5M_\odot$ and a projected separation of $\sim$2000 AU \citep{Mugrauer2014HD4113B, Stassun2019hd4113B}.


We aim to reproduce the observationally constrained system parameters of HD 4113 through secular chaos. To directly check for instabilities in the system, we use the \texttt{REBOUND} numerical integration package to run N-body simulations \citep{rein2012rebound}. We use REBOUNDx to include general relativity and tidal forces \citep{tamayo2019reboundx}. General relativity is included through the \texttt{gr\_full} prescription \citep{Newhall1983grfull}. Equilibrium tides are incorporated with the \texttt{tides\_spin} prescription \citep{eggleton1998tidespin, lu2023tidespin}, and dynamical tides are incorporated with the \texttt{tides\_dynamical} prescription \citep{Vick2019dynmtides, Liveoak2025dynmtides}. The parameters used for the simulation that best reproduced the observed system parameters are listed in Table \ref{tab:hd4113}. To ensure that $\mathcal{R}$ is sufficiently large for secular chaos, we adopt an outer eccentricity of $e_3\simeq 0.85$. This choice is reasonable for wide binary orbits, whose eccentricity distribution is observed to be superthermal and becomes increasingly weighted toward large eccentricities at wider separations \citep{Tokovinin2016eccdist, Hwang2022eccdist}. Because wide binaries are most likely observed near apoapsis, we set $a_3$ such that $a_3(1+e_3)\approx 2000$ AU, yielding $a_3\approx 1100$ AU.

\begin{figure}
    \centering
    \includegraphics[width=0.98\linewidth]{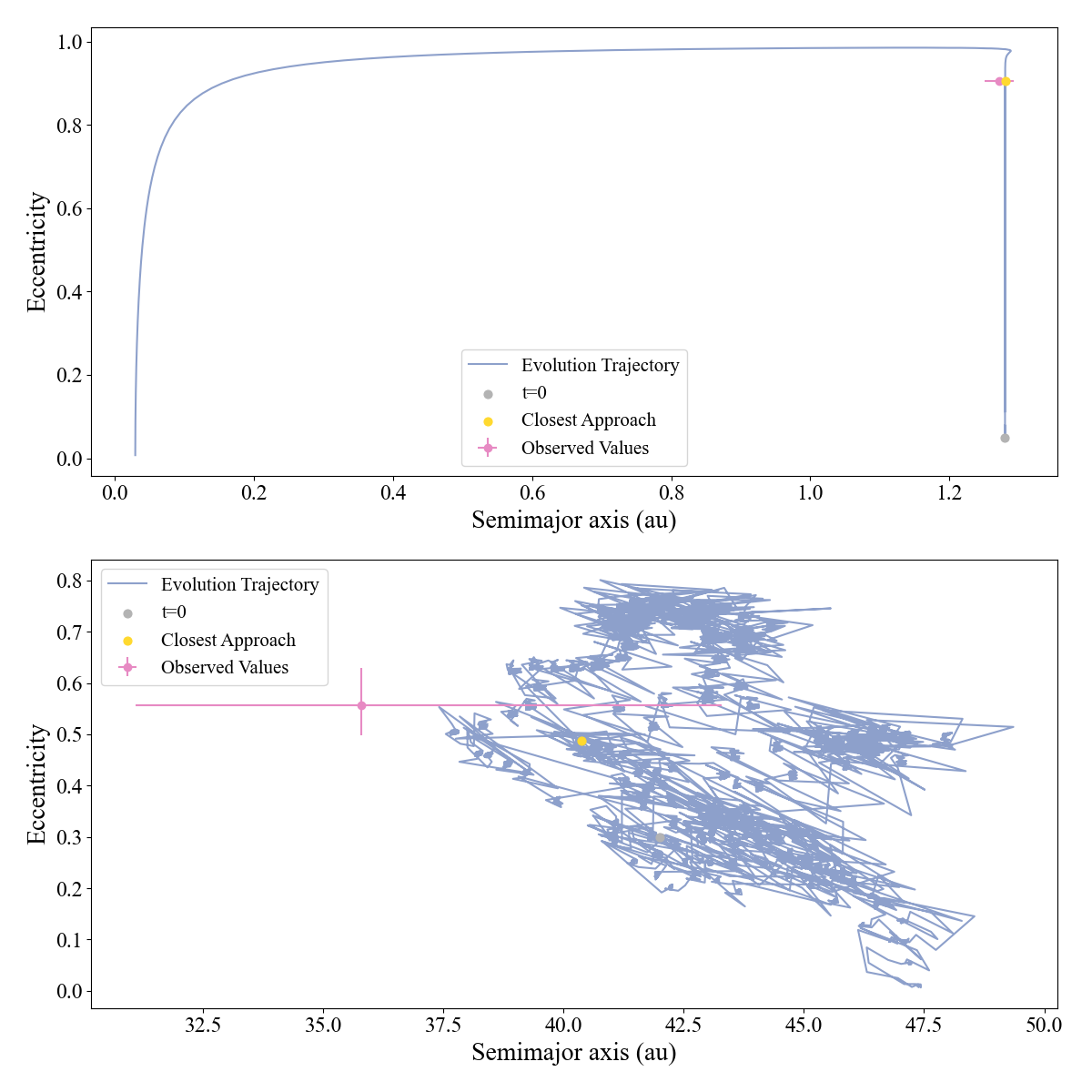}
    \caption{The eccentricity and semimajor axis evolution of a simulation of HD 4113 b and HD 4113 C that matches the observed values. The system parameters are shown in Table \ref{tab:hd4113}.}
    \label{fig:application}
\end{figure}

The results of the simulation in the $e_1$-$a_1$ plane and the $e_2$-$a_2$ plane are shown in Figure \ref{fig:application}. We start the simulation with $e_1=0.05$, and it is excited to its current observed value through secular chaos. Though the orbital evolution of the planet still follows the prediction from secular chaos, we observe stochastic behavior in $e_2$-$a_2$ space due to N-body effects. Because of this stochastic behavior, the value of $\mathcal{R}$ varies from 0.15 to 1.5, with a time-average before circularization of 0.64, whereas in the corresponding secular simulation, the value of $\mathcal{R}$ varies from 0.48 to 0.75, with a time-average before circularization of 0.61. In the N-body simulation, the larger variance in $\mathcal{R}$ has caused the planetary eccentricity to be excited to near-unity more rapidly. We check that the planet avoids crossing the Roche limit \citep{matsumura_2010_tidal, naoz2012formation, Petrovich2015}.


\paragraph{HAT-P-7 and TOI-179}
Other ``3+1'' systems, including HAT-P-7 \citep{Winn2009hatp7, yang2025hatp7} and TOI-179 \citep{Desidera2023toi179}, are more difficult to explain with secular chaos because the possible $\mathcal{R}$ values for these systems are restricted to a low range by their known system parameters ($\mathcal{R}<10^{-2}$). For secular chaos to be effective, either $\mathcal{R}$ has to be close to unity or $I_{23}$ has to be high, as shown in Figure \ref{fig:increaseinc}. For these systems, only extreme $I_{23}$ values can allow for secular chaos to happen.

\subsubsection{Where are the hidden low-mass companions?}
\label{sec:hidden}

Roughly half of solar-type stars host binary companions \citep{Raghavan2010}.  It has been found that approximately a quarter of exosystems reside in multiple-star systems \citep{Fontanive2021}, most commonly with wide-orbit companions ($a \gtrsim 100$ AU) \citep[e.g.,][]{MoeKratter2021,Lester2021}.  In addition, both astrometric and high-resolution imaging surveys indicate that wide binaries are generally aligned with close-in planetary systems \citep{christian2022possible,dupuy2022orbital,lester2023visual}.  These findings are consistent with the model presented in this work. 
However, to date, only three ``3+1'' systems have been identified with an intermediate companion. This naturally raises the question: where are the low-mass intermediate companions?

Disk fragmentation can produce wide orbit brown dwarf companions \citep{Stamatellos2009} and Jupiters \citep{Vorobyov2018}. While some studies suggest that a nearby stellar companion may suppress planet formation \citep{Kraus2012}, a wider companion appears to have negligible influence \citep{MoeKratter2021}. Furthermore, recent SPH simulations demonstrate that intermediately separated stellar companions  can trigger fragmentation in protoplanetary disks at tens of AU \citep{Cadman2022}. 
It is therefore expected that many giant planets and substellar companions remain to be discovered.

Although detecting cold Jupiters is challenging with current observations, the upcoming release of Gaia DR 4 is expected to significantly expand the number of detected cold Jupiters, possibly discovering  ``3+1'' systems with cold Jupiter perturbers. From Figure 13 of \cite{Wallace2025gaia} and similar Gaia DR 4 sensitivity function figures, we can see that planets of mass ratios of around $m_2/m_0 =0.01$ ($\sim10 M_{\mathrm{Jup}}/1M_\odot$) can be detected up to 100 year periods for nearby stars, which translates to $\sim 20$ AU around a sun-like star. Therefore, cold Jupiter perturbers suitable for inducing secular chaos in ``3+1'' systems should be marginally detectable with Gaia DR 4. Gaia DR 5 will further expand the detectable parameter space. Brown dwarf perturbers are more massive and thus are easier to detect with Gaia, so we may also be able to discover more ``3+1'' systems with brown dwarf perturbers as well.

\subsubsection{Degeneracy between system spacing and tidal dissipation efficiency}\label{sec:degeneracy}


For fixed $\mathcal{R}$, the tidal-disruption boundary reflects a degeneracy between the planetary viscous timescale and the system spacing. Weaker tidal dissipation, corresponding to larger $t_{V,1}$, can be accommodated only in more widely separated systems, where secular forcing is slower. Conversely, in compact configurations, survival requires sufficiently small $t_{V,1}$ so that tidal damping can suppress eccentricity growth before disruption occurs.

For the fiducial setup in Table \ref{tab:fiducial}, avoiding tidal disruption requires $t_{V,1} \lesssim 10^{-3}$ yr. Increasing the system scale relaxes this constraint. For example, when $a_2$ is increased to 70 AU, with $a_3$ adjusted to keep $\mathcal{R}=1$, the survival threshold becomes $t_{V,1} \lesssim 4.7\times 10^{-3}$ yr. For $a_2=110$ AU, it rises further to $t_{V,1} \lesssim 0.1$ yr. The resulting disruption boundary is approximately linear in the $\log_{10} t_{V,1}$--$a_2$ plane. This trend arises from the competition between secular excitation, which drives eccentricity growth toward disruption, and tidal damping, which counteracts that growth. As the system spacing increases, the secular forcing timescale lengthens, allowing even weaker tidal dissipation to damp the orbit rapidly enough to prevent disruption.

In this work, we use $t_{V,1} = 10^{-3}$ in secular simulations, following the choice of $t_{V,1}$ in \cite{Fabrycky2007ZLK}. This choice of viscous timescale is equivalent to a Q of $\sim10$ when the forcing frequency is set as the mean motion\footnote{The estimated value of Q is calculated using the relationship between Q and $t_{V,1}$ \citep[e.g.][]{eggleton1998tidespin, Eggleton2001,fabrycky2007shrinking}: $Q=\frac{4}{3}\frac{k_{2,1}}{(1+2k_{2,1})^2}\frac{Gm_1}{R_1^3}\frac{t_{v,1}}{n_1}$}. Such strong tidal dissipation is plausible in the high eccentricity migration regime, where diffusive excitation of the planetary f-modes and r-modes during repeated pericenter passages leads to non-linear dissipation that rapidly shrinks the planet's orbit \citep{vick_2019}. This process can circularize a Jovian planet on a timescale of only ~$10^4$ years, corresponding to Q$\sim$1 \citep{Wu2018lowQ}. Other studies have adopted $t_{V,1} = 0.1$ years  \citep{Petrovich2015} and $t_{V,1}=1.5$ years \citep{naoz2012formation} as the fiducial value.

Lastly, we note that the zone of tidal disruption used in this work is $a_1(1-e_1)\leq 2.7r_t$. Here, $r_t=r_1(m_1/(M_0+m_1))^{-1/3}$ is the characteristic tidal radius \citep[e.g.][]{Petrovich2015,naoz2016eccentric}. However, recent work using simulations of giant planets with realistic dense cores has shown that no tidal disruption occurs within this zone, and only in the region of $a_1(1-e_1)\leq2.0r_t$ does significant mass loss occur \citep{Fan2026TidalDisruption}. Therefore, crossing the tidal disruption radius of $\leq2.7r_t$ used in this work does not guaranty planetary destruction. Instead, this criterion should be interpreted as a conservative marker for potentially tidal disruption.

\subsection{Obliquity distribution}
\label{sec:psi_dist}


\subsubsection{Comparison with observations}
The secular chaos mechanism in ``3+1'' systems, introduced in this work, typically produces final stellar obliquities slightly below $90^\circ$. Notably, this outcome is insensitive to the value of $\mathcal{R}$. The excess of polar planets seen in our simulations resembles observational trends that have been reported.

Earlier studies suggested a prevalence of perpendicular planets \citep[e.g.][]{Albrecht2021perpendicular}, but more recent work indicates that most planets are aligned, with misaligned systems exhibiting an isotropic obliquity distribution \citep{Siegel2023perpendicular, dong2023Align, Rossi2025align}. Restricting attention to sub-Saturns and hot Jupiters around F-type stars, tentative evidence points to a peak near $90^\circ$ \citep{Knudstrup2024obliquity, Dugan2025}. Similarly, the current sample of hot Neptunes shows a possible clustering of polar orbits \citep{Espinoza2024neptune}, though larger samples are required to confirm this trend \citep{Rossi2025align}. While our analysis focuses on Jupiters, the mechanism applies broadly to companions across the planetary mass spectrum.

Meanwhile, the correlation between the presence of a stellar companion and spin-orbit misalignment is ambiguous. \cite{Ngo2015} finds no correlation between misaligned hot Jupiters and the presence of wide stellar companions. Meanwhile, \citet{EspinozaRetamal2023} find that eccentric warm and hot Jupiters in binary systems tend to exhibit large obliquities, although the sample remains small.

Looking ahead, as more “3+1’’ systems are identified, it may become possible to test whether binary systems with intermediate companions are more likely to host perpendicular planets than those without, specifically within the parameter space where “3+1’’ secular chaos operates effectively.

\subsubsection{Alternatives to secular chaos to produce nearly polar planets}

In addition to secular chaos, another pathway capable of naturally producing polar orbits is disk-driven resonance, which can generate close-in sub-Neptunes on polar trajectories in the presence of outer cold Jupiter companions \citep{Petrovich2020neptune, Louden2024neptune}. By contrast, pure ZLK migration does not typically yield strictly perpendicular planets; instead, it produces a bimodal obliquity distribution \citep{Fabrycky2007ZLK}. An exception arises when a primordial, disk-induced spin--orbit misalignment is present: under such circumstances, the resulting stellar obliquity exhibits a broad peak near $90^\circ$ \citep{Vick2023Perpendicular}.

A planet may also acquire a polar obliquity \emph{primordially}. A misalignment between the stellar spin axis and the protoplanetary disc can originate from several processes, including magnetic star--disc interactions \citep{Lai2011EvolSpin, Spalding2015}, torques induced by binary companions \citep{Batygin2012, Lai2014}, and turbulence within the disc \citep{Fielding2015}. Recent observations further suggest that a significant fraction of young stars may form within misaligned, planet-forming discs \citep{Biddle2025}.

Although tidal dissipation is generally expected to damp obliquities, certain conditions allow planets to retain large stellar obliquities. \citet{Lai2012} showed that in misaligned systems, inertial waves excited within the convective envelopes of cool, solar-type stars—driven by the Coriolis force—can establish new spin equilibria that support long-lived polar obliquities. In contrast, hot stars lack substantial convective envelopes, rendering this mechanism ineffective in those systems.

\subsection{Comparison with eccentricity cascade from \cite{yang2025hatp7}}
\label{sec:compYang2025}

Another pathway for driving eccentricity growth in “3+1” systems is the eccentricity cascade mechanism described by \cite{yang2025hatp7}. In this scenario, the brown dwarf’s eccentricity is first excited through ZLK oscillations induced by the distant stellar companion, and this excitation is subsequently transferred to the planet through close encounters. This mechanism differs from the one explored in this work in several fundamental ways:

\begin{itemize}
\item The eccentricity cascade relies on non-secular perturbations to transfer eccentricity, whereas our mechanism operates entirely within the secular regime, with no direct energy exchange between orbits.
\item Eccentricity growth in the cascade requires the mutual inclination between the brown dwarf and the distant stellar perturber to be high enough to trigger ZLK oscillations. In contrast, the mechanism studied here remains effective even at low mutual inclinations; as shown in Figure~\ref{fig:increaseinc}, secular chaos can operate for inclinations below $20^\circ$.
\item In the eccentricity cascade, the planet and brown dwarf remain tightly coupled, with low mutual inclinations $I_{12}$, whereas in our scenario, the system exhibits chaotic evolution of $I_{12}$, leading to qualitatively different dynamical behavior.
\end{itemize}


To draw a direct comparison with Figure 8 in \cite{yang2025hatp7}, we create Figure \ref{fig:fractioninc}, where we show the fraction of the three outcomes—unexcited, excited, and unstable per the \cite{Tory2022instability} criterion—as a function of the initial $I_{23}$. Compared to Figure 8 in \cite{yang2025hatp7}, where the fraction of excited systems is below 0.2 for initial mutual inclinations of $\sim40^\circ$, the mechanism in this work excites all systems with initial $I_{23}\sim40^\circ$ under $\mathcal{R}=1$. This shows that eccentricity excitation through secular chaos in ``3+1'' systems is significantly more effective in the low inclination regime.

The two mechanisms operate under different conditions and are complementary in nature. Each tends to produce a different final system architecture—for example, coupled versus uncoupled inner bodies. At present, the number of known “3+1” systems is too small to assess how relevant these mechanisms are to real hot-Jupiter–hosting systems. If enough ``3+1'' hot Jupiter-hosting systems are detected with Gaia DR 4 and long-term radial velocity campaigns, we may begin to draw population-wide conclusions about how relevant these ``3+1'' mechanisms are to the formation of hot Jupiters in binary systems by comparing the predicted system architecture outcomes of these mechanisms to observables such as the occurrence rate of cold Jupiters in hot Jupiter-hosting binaries and the mutual inclination distributions \citep[e.g.,][]{EspinozaRetamal2023}. For example, if in such ``3+1'' systems, the mutual inclination between the intermediate perturber and the stellar companion is distributed predominantly below $40^\circ$, similar to $I_{23}$ in panel (c) of Figure \ref{fig:obliquity}, then the mechanism described in this work is preferred. Conversely, if $I_{12}$ is clustered around $0^\circ$, then the eccentricity cascade mechanism is preferred.

\section{Conclusions}
\label{sec:conclusion}

In this work, we study the conditions under which secular chaos can trigger high-eccentricity tidal migration in “3+1” systems—stellar binaries hosting a planet and an additional companion orbiting one of the stars. We show that chaotic secular evolution is controlled by the ratio of timescales $\mathcal{R}=t_{\mathrm{ZLK},12}/t_{\mathrm{ZLK},23}$ in Equation (\ref{eq:R}). Values of $\mathcal{R}$ near unity (within a factor of a few) enable high-eccentricity migration across most initial conditions, even when mutual inclinations lie below the ZLK critical angle.

Other key results for the four-body dynamics include:
\begin{itemize}
\item Using surfaces of section constructed from a ZLK-like Hamiltonian in a frame rotating under the influence of the wide binary, we demonstrate that “3+1” systems exhibit broad chaotic regions and narrow libration islands when $\mathcal{R}\sim 1$.
\item We find that although the time required to drive the planet to near-unity eccentricity increases with $\mathcal{R}$, the fraction of systems that achieve such excitation rises at low $\mathcal{R}$, plateaus around $\mathcal{R}\approx 0.6$–1.0, and declines for larger values.
\item When the low–mutual-inclination condition is relaxed, planets are more easily excited to near-unity eccentricities and over a wider range of $\mathcal{R}$, but the resulting architectures are more prone to dynamical instability.
\item We constrain the system parameters of an observed ``3+1'' system, HD 4113 using archival data, and apply this mechanism to explain the system's architecture, demonstrating its applicability to a real system.
\end{itemize}

Our population synthesis shows that migrated planets (hot Jupiters or hot Neptunes) settle into stellar obliquities centered near $\sim 80^\circ$, and highly tilted orbits relative to the outer companions (median mutual inclinations of $\sim 70^\circ$). Future Gaia releases (DR 4 and DR 5) and  long-term radial velocity campaigns may uncover additional “3+1” systems and provide inclination constraints, offering new opportunities to test these theoretical predictions.


\begin{acknowledgments}
Y.L. is funded by the Alexander P. Hixon Fellowship. Y.L. gratefully acknowledges the Alice Palma Summer Undergraduate Research Program for hosting her at Indiana University Bloomington. Y.L. thanks Yubo Su, Malena Rice, and Tiger Lu for helpful discussions. C.P. is supported by the National Science Foundation under Grant No. AST-2511257.

\end{acknowledgments}

\appendix

\section{Stellar Mass and Orbital Constraints on HD\,4113}
\label{sec:hd4113_orbit}

To provide physically motivated initial conditions for our dynamical integrations of HD\,4113, we performed a joint radial-velocity and astrometric fit to the two known companions in the system. The fit combines literature radial velocities, direct-imaging astrometry of HD\,4113\,C, and the Hipparcos--Gaia proper-motion anomaly \citep[e.g.,][]{Tamuz2008HD4113,Cheetham2018hd4113,Feng2022hd4113}.

\subsection{Stellar Mass Priors}

The joint RV+astrometry analysis requires a prior on the stellar mass. To derive this prior, we performed spectral energy distribution (SED) modeling with \texttt{EXOFASTv2} \citep{Eastman2017, Eastman2019}, combined with MESA Isochrones and Stellar Tracks (MIST; \citealt{Choi2016mist, Dotter2016mist}). The broadband photometry used in the SED fit included Gaia $G$, $G_{\rm BP}$, and $G_{\rm RP}$; 2MASS $J$, $H$, and $K$; and WISE $W1$, $W2$, and $W3$ measurements. We adopted a Gaussian prior on metallicity (\feh) from \citet{Sousa2021}\footnote{\url{https://sweetcat.iastro.pt/}}, $\feh=\mathcal{G}(0.23, 0.08)$, using a conservative 0.08 dex uncertainty instead of the reported 0.01 dex value. A Gaussian prior on the parallax was adopted from Gaia DR3 \citep{GaiaCollaboration2023}, after applying the zero-point correction from \citet{Lindegren2021}. The upper limit on the $V$-band extinction ($A_V$) was set based on \citet{Schlafly2011}.

The posterior uncertainties were estimated using a parallel-tempering Differential Evolution Markov chain Monte Carlo algorithm (PT-DE-MCMC; \citealt{braak2006markov}). The number of temperatures was set to eight, and the number of walkers was automatically set by \texttt{EXOFASTv2} to twice the number of free parameters, yielding 22 walkers. We considered the chains converged when the number of independent draws, $T_z$, exceeded 1000 and the Gelman-Rubin statistic, $\hat{R}$, was less than 1.01. The resulting stellar mass is $0.976^{+0.046}_{-0.035}\,\msun$. To account for systematic uncertainties among different stellar-evolution models, we adopted a conservative 5\% uncertainty. Thus, the stellar mass of HD\,4113 adopted in the following analysis is $0.976\pm0.049\,\msun$.

Our dynamical simulations also require the stellar mass of HD\,4113\,B. Because HD\,4113\,A and B form a wide bound binary, we assumed common priors on \feh, parallax, and $A_V$, and applied the same SED+MIST modeling procedure to HD\,4113\,B. After adopting the same conservative 5\% uncertainty, we obtained a stellar mass of $0.556 \pm 0.028\,\msun$.

\begin{figure*}
    \centering
    \includegraphics[width=1\linewidth]{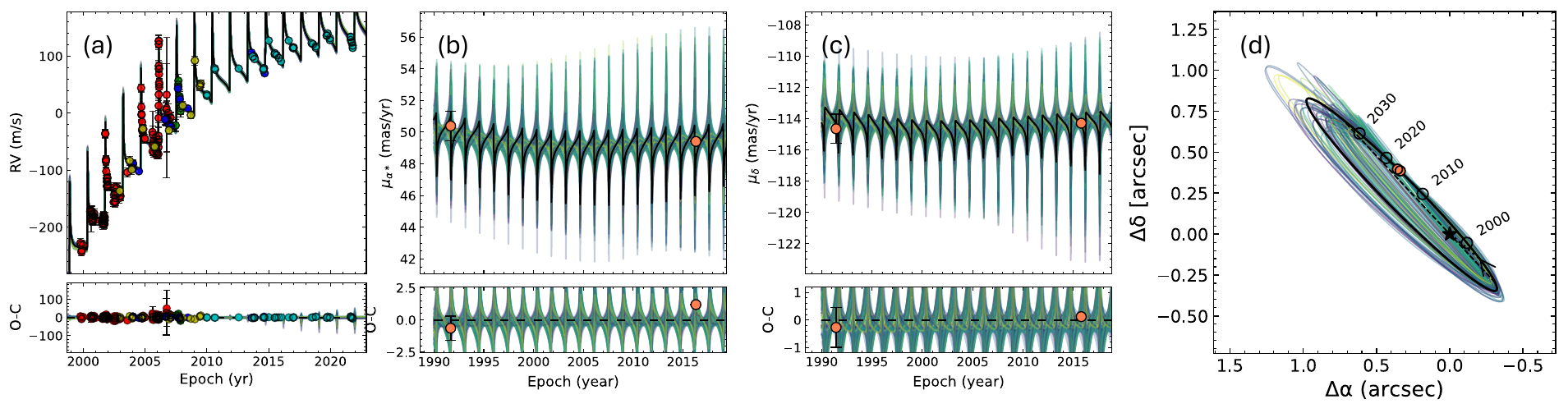}
\caption{Joint two-companion orbit fit of HD\,4113 with \texttt{orvara}.
\textbf{Panel (a):} Radial-velocity time series with the best-fit model overlaid,
colored by instrument (red: CORALIE-98, before the 2007 upgrade; green: CORALIE-07, after the 2007 upgrade; blue: HIRES; yellow: MIKE; cyan: PFS); the lower subpanel shows the residuals.
\textbf{Panels (b) and (c):} Hipparcos--Gaia proper motions in
$\mu_{\alpha*}$ and $\mu_{\delta}$, respectively. The dense teal curves show 100
random posterior orbit realizations, the orange points show the three
catalog-derived proper-motion measurements (Hipparcos, Hipparcos--Gaia scaled,
and Gaia~EDR3), and the lower subpanels show the corresponding residuals.
\textbf{Panel (d):} Posterior orbital trace of HD\,4113\,C relative to the host star
on the sky, with the two SPHERE astrometric epochs shown in orange.}
    \label{fig:hd4113_orvara}
\end{figure*}

\subsection{Joint RV and Astrometry Modeling}

To constrain the orbit of the HD\,4113 system, we performed a joint fit of 203 radial velocities from CORALIE \citep{CORALIE} on the 1.2\,m Swiss Euler telescope, HIRES \citep{HIRES} on the Keck 10\,m telescope, and MIKE \citep{MIKE} and PFS \citep{crane2006carnegie,crane2008carnegie,crane2010carnegie} on the Magellan Clay 6.5\,m telescope. We also included two SPHERE \citep{SPHERE2013} relative-astrometry epochs from \citet{Cheetham2018} and the Hipparcos--Gaia astrometric acceleration from the Hipparcos--Gaia Catalog of Accelerations \citep[HGCA;][]{HGCA}. The joint fit was performed using the publicly available \texttt{orvara} package \citep{orvara}.

We placed a Gaussian prior on the primary mass, $M_\star = 0.976 \pm 0.049\,M_\odot$, derived from our SED+MIST modeling. The orbit of HD\,4113\,A\,b was further constrained by a Gaussian prior on the orbital period, $P_b = 526.0 \pm 0.5$\,d, motivated by the discovery RV solution of \citet{Tamuz2008HD4113}, and by a uniform eccentricity prior, $e_b \sim \mathcal{U}(0.85, 0.95)$, which brackets the high-eccentricity solution established by earlier studies. For HD\,4113\,C, we adopted broad priors, with $e_C \sim \mathcal{U}(0,0.9)$ and the remaining Keplerian elements drawn from the default \texttt{orvara} ranges. An independent jitter term, sampled uniformly from $[10^{-5}, 10^{3}]$\,m\,s$^{-1}$, was assigned to each of the five RV instruments.

Posterior sampling used the parallel-tempered ensemble sampler \citep[\texttt{ptemcee};][]{ptemcee}, with eight temperatures, 64 walkers, and 500,000 steps per walker. The Gelman--Rubin statistic was $\hat{R} \le 1.01$ for all 15 free orbital parameters.

For HD\,4113\,C, all orbital parameters agree with \citet{Feng2022hd4113} to within $3\sigma$, except for the argument of periastron, which differs by $3.5\sigma$. This residual tension likely reflects differences between the sampling and modeling algorithms used by \texttt{orvara} and those adopted by \citet{Feng2022hd4113}, as well as the different stellar-mass priors. In this work, we adopt $M_\star = 0.976 \pm 0.049\,M_\odot$, whereas \citet{Feng2022hd4113} adopted the TESS Input Catalog value, $M_\star = 1.00 \pm 0.12\,M_\odot$ \citep{TICInputCatalog}. Future Gaia DR4 and DR5 epoch astrometry will further refine the orbit of HD\,4113\,C and help resolve this remaining discrepancy.

\bibliography{sample7}{}
\bibliographystyle{aasjournalv7}

\end{document}